\documentclass[aps,prl,showpacs,twocolumn,superscriptaddress]{revtex4}
\usepackage{bm,color}
\usepackage{graphicx}
\usepackage{amsmath, amssymb}
\usepackage{braket}
\usepackage{ulem}
\begin{document}
\title{Reservoir Computing with Spin Waves in Skyrmion Crystal}
\author{Mu-Kun Lee}
\affiliation
{Department of Applied Physics, Waseda University, Okubo, Shinjuku-ku, Tokyo 169-8555, Japan}
\author{Masahito Mochizuki}
\affiliation
{Department of Applied Physics, Waseda University, Okubo, Shinjuku-ku, Tokyo 169-8555, Japan}
\begin{abstract}
Magnetic skyrmions are nanometric spin textures characterized by a quantized topological invariant in magnets and often emerge in a crystallized form called skyrmion crystal in an external magnetic field. We propose that magnets hosting a skyrmion crystal possess high potential for application to reservoir computing, which is one of the most successful information processing techniques inspired by functions of human brains. Our skyrmion-based reservoir exploits precession dynamics of magnetizations, i.e., spin waves, propagating in the skyrmion crystal. Because of complex interferences and slow relaxations of the spin-wave dynamics, the skyrmion spin-wave reservoir attains several important characteristics required for reservoir computing, e.g., the generalization ability, the nonlinearity, and the short-term memory. We investigate these properties by imposing three standard tasks to test the performances of reservoir, i.e., the duration-estimation task, the short-term memory task, and the parity-check task. Through these investigations, we demonstrate that magnetic skyrmion crystals are promising materials for spintronics reservoir devices. Because magnetic skyrmions emerge spontaneously in magnets via self-organization process under application of a static magnetic field, the proposed skyrmion reservoir requires neither advanced nanofabrication nor complicated manufacturing for production in contrast to other previously proposed magnetic reservoirs constructed with fabricated spintronics devices. Our proposal is expected to realize a breakthrough in the research of spintronics reservoirs of high performance.
\end{abstract}
\maketitle
%\sloppy

\section{Introduction}
%%%%%%%%%%%%%%%%%%%%%%%%%%%%%%%%%%%%%%%%%%%%%%%%%%%%%%%%%%%%%
\begin{figure}[tb]
\centering
\includegraphics[scale=0.5]{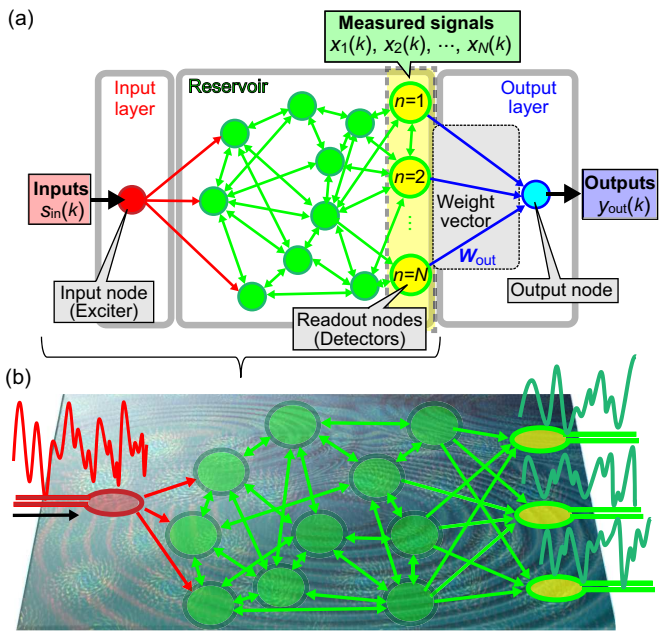}
\caption{(a) Architecture of the reservoir computing. The weight vector ${\bm W}_{\rm out}$ connecting the reservoir and the output node is required to be optimized by training to produce desired outputs for a given problem. (b) Proposed physical reservoir exploiting spin waves propagating in a thin-plate magnet hosting a magnetic skyrmion crystal. Single or multiple microwave generators are installed near the left edge as input nodes where locally applied microwave fields corresponding to the input data induce magnetization oscillations in the skyrmion crystal. Single or multiple readout nodes called detectors are installed near the right edge which measure local dynamics of magnetizations via an electric voltage induced by the electromagnetic induction. The signals measured at the detectors are translated to outputs via a linear transformation using the optimized weight matrix ${\bm W}_{\rm out}$.}
\label{Fig01}
\end{figure}
%%%%%%%%%%%%%%%%%%%%%%%%%%%%%%%%%%%%%%%%%%%%%%%%%%%%%%%%%%%%%
Reservoir computing~\cite{Tanaka2019,Nakajima2020,Nakajima2018} is one of the most successful information processing techniques inspired by the nerve system of human brains, which is composed of three sectors, i.e., input layer, reservoir, and output layer [see Fig.~\ref{Fig01}(a)]. Sequential information data are entered via the input layer composed of one or more input nodes. The input signals are transformed to different signals in a nonlinear way through recursively propagating in a dynamical medium called reservoir. Through this nonlinear transformation, the input data are mapped onto a higher-dimensional information space. Reservoirs are composed of many ingredients connected mutually via physical interactions~\cite{Tanaka2019,Nakajima2020} to achieve the nonlinear transformation and are used as a black box in computing because the connections among the ingredients are complicated and usually unknown. The transformed signals are measured at installed readout devices and are, subsequently, translated to outputs via a linear transformation with a weight vector ${\bm W}_{\rm out}$. The weight vector ${\bm W}_{\rm out}$ connects the readout nodes and the output node, which is optimized by training so as to provide correct answers or desired outputs for a given problem. In the reservoir computing, we only need to train the weight vector ${\bm W}_{\rm out}$ in contrast to another established brain-inspired information processing technique called neural network computing~\cite{Williams1989,Werbos1990}, in which all the weight matrices $\bm W_{\rm res}$ connecting nodes in the multiple layers must be optimized.

A lot of physical reservoirs have been proposed and demonstrated to date, which are based on, e.g., optical~\cite{Optical01,Optical02,Optical03,Optical04}, mechanical~\cite{Mechanical01,Mechanical02,Mechanical03,Mechanical04}, biological~\cite{Biological01,Biological02,Biological03,Biological04}, electronic~\cite{Electronic01,Electronic02,Electronic03,Electronic04}, and magnetic systems~\cite{Magnetic01,Kanao2019,STO02,STO03,Furuta2018, Nakane2018,Yamaguchi2020, Bourianoff2018, Prychynenko2018, Pinna2020,Jiang2019}. Among them, the magnetic reservoirs have attracted a great deal of research interest because they have numerous advantages over other physical reservoirs. The advantages of magnetic reservoirs are (1) nonvolatility~\cite{Grollier2016}, (2) durability against environmental agitations~\cite{YuXZ2011,Zhang2014}, e.g., radiations, heats, and mechanical shocks, (3) low energy consumption~\cite{Joshi2016}, and (4) quick responses~\cite{Kanao2019}. The former two advantages are pronounced in comparison with reservoirs based on the optical systems and semiconductors, while the latter two advantages are pronounced in comparison with the mechanical reservoirs.

Most of the previously proposed magnetic reservoirs are based on series-connected spin-torque oscillators~\cite{Magnetic01,Kanao2019,STO02,STO03,Furuta2018}. A spin-torque oscillator is a micrometric or even nanometric sized spintronics device that has a layered nanopillar structure with soft and hard ferromagnetic layers separated by a thin insulating layer. The magnetization in the soft ferromagnetic layer exhibits specific resonant oscillations upon injection of spin-polarized currents or application of microwaves, while the magnetization in the hard ferromagnetic layer is fixed and hardly changes its orientation. The relative magnetization directions between the soft and hard ferromagnetic layers affect the electric resistance of currents tunneling through the thin insulating layer, and thus the magnetization oscillations in the soft ferromagnetic layer can be detected by the measurement of electric resistance. Spin-torque oscillators in a reservoir interact magnetically via magnetic dipole-dipole interactions or electrically via circuit connections to realize the reservoir functions. Several numerical simulations~\cite{Kanao2019,Furuta2018} have demonstrated that the spin-torque oscillator reservoirs show high performances in information processing. However, it is required to fabricate a number of high-quality spin-torque oscillators of nanometric size with uniform characteristics to realize the reservoir functions. Thereby, their productions require advanced nanofabrication techniques, complicated manufacturing processes, and high costs. Under these circumstances, high-performance magnetic reservoirs, which can be produced easily with low costs, are demanded.

Recently, nanometric topological spin textures called skyrmions are attracting a great deal of research interest~\cite{Pfleiderer2011,Nagaosa2013,Fert2013,Seki2015,Everschor-Sitte2018}. The concept of skyrmion was originally proposed in 1960s as a mathematical model of baryon in particle physics~\cite{Skyrme1961,Skyrme1962}, which comprises vector fields pointing in all directions wrapping a sphere like a hedgehog. In 1980s, it was theoretically predicted that skyrmions can emerge in ferromagnets with broken spatial inversion symmetry as a two-dimensional vortex-like spin texture, which can be regarded as a stereo projection of the original hedgehog-type skyrmion onto a plane~\cite{Bogdanov1989,Bogdanov1994,Rossler2006}. They also predicted that magnetic skyrmions often emerge in a hexagonally crystallized form called skyrmion crystal. In 2009, the magnetic skyrmion crystals were indeed discovered in metallic chiral magnets by small-angle neutron scattering measurements~\cite{Muhlbauer2009} and Lorentz transmission electron microscopies~\cite{YuXZ2010}. Skyrmions are characterized by a quantized topological invariant called skyrmion number representing how many times the magnetizations wrap a sphere~\cite{Nagaosa2013,Braun2012}. This means that magnetic skyrmions belong to a different topological class from ferromagnetic and conical/helical states, and, thereby, we cannot create and annihilate them in a uniform ferromagnetic state by continuous variation of the spatial alignment of magnetizations~\cite{Braun2012}. Instead, a local reversal of magnetization is required for their creation and annihilation, which necessarily costs a large energy associated with the ferromagnetic exchange interaction. Owing to the protection by this energy cost, the magnetic skyrmions are robust against environmental agitations and external stimuli~\cite{YuXZ2011}.

It was theoretically revealed that a magnetic skyrmion crystal exhibits peculiar spin-wave modes at microwave frequencies, i.e., clockwise and counterclockwise rotation modes active to in-plane microwave fields and a single breathing mode active to out-of-plane microwave fields~\cite{Mochizuki2012,Petrova2011,Mochizuki2015}. Magnetic skyrmions constituting a skyrmion crystal uniformly rotate in the two rotation modes, whereas they uniformly expand and shrink in an oscillatory manner in the breathing mode. These modes are similar to the collective modes of a magnetic vortex confined in a magnetic nanodisk or nanopillar~\cite{Nanodisk01,Nanodisk02}. In this sense, each skyrmion in a skyrmion crystal can be regarded as a spin-oscillator device carrying the eigenmode spin oscillations. This fact indicates that a magnetic skyrmion crystal and even an assembly of randomly aligned magnetic skyrmions can work as series-connected spin-torque oscillators. It further comes up with an idea of magnetic reservoirs exploiting the spin waves or collective magnetization dynamics of magnetic skyrmions. One advantage of the usage of magnetic skyrmions for reservoirs is that neither advanced nanofabrication techniques nor complicated manufacturing processes are required for production in contrast to the previously proposed and widely investigated spin-torque-oscillator reservoirs because magnetic skyrmions can be created simply by application of static magnetic field to a plate-shaped sample of chiral magnet or a magnetic bilayer system~\cite{YuXZ2010}.

In this paper, we propose a new magnetic reservoir device, which exploits spin waves or magnetization dynamics propagating in a plate-shaped magnet hosting magnetic skyrmions. We demonstrate that this skyrmion spin-wave reservoir possesses several important characteristics required for reservoir computing, e.g., the generalization ability~\cite{Nakane2018}, the nonlinearity~\cite{Furuta2018,Kanao2019}, and the short-term memory~\cite{Furuta2018,Kanao2019} owing to their complex interferences and slow relaxations, by imposing three standard tasks to test the performances, i.e., the duration-estimation task, the short-term memory task, and the parity-check task~\cite{Furuta2018,Kanao2019}. Through these investigations, we argue that magnetic skyrmions are promising materials for spintronics reservoirs. Because the proposed skyrmion spin-wave reservoir has a lot of advantages over other previously proposed magnetic reservoirs consisting of fabricated spintronics devices, the present work will necessarily mark a great progress in the research on the reservoir computing.

\section{Concept and Method}
\subsection{Skyrmion Spin-Wave Reservoir}
We first discuss the concept of our skyrmion spin-wave reservoir as shown in Fig.~\ref{Fig01}(b). A key ingredient of this reservoir is a thin-plate specimen of skyrmion-hosting magnet. 
%More specifically, the specimen hosts an assembly of magnetic skyrmions in which the skyrmions are aligned to form a deformed triangular lattice or even in a random configuration. 
One or more devices called input nodes are fabricated on the left side of the specimen to enter signals as input data. The entered signals are transformed while they propagate in the specimen recursively. Other devices called readout nodes or detectors are fabricated on the right side of the specimen to measure the transformed signals.

The information processing with this skyrmion-based reservoir system is performed as follows. We apply out-of-plane microwave magnetic fields $\bm H^\omega$ to enter the input data, which are locally generated by injecting electric currents $j_{\rm in}$ to a metallic ring fabricated as an input node on the specimen. More concretely, a series of input data are entered after being translated to time-dependent amplitude and duration of the microwave pulses. The applied microwave pulses induce magnetization oscillations in the skyrmions, and the induced oscillations propagate in the specimen as spin waves, which exhibit nonlinear or even chaotic behaviors because of complicated interferences due to the distorted configuration of hexagonal skyrmion crystal. The magnetization oscillations eventually reach the area of readout nodes (detectors). The magnetization dynamics within an area of each detector are measured as time-profiles of induced electric currents via the electromagnetic induction.

This skyrmion spin-wave reservoir has advantages over the magnetic reservoirs based on the spin-torque oscillators. First, complicated nanofabrication is not required because skyrmions are created spontaneously in magnets by application of magnetic field~\cite{YuXZ2010}. We can prepare a skyrmion crystal or an assembly of skyrmions by simply applying a magnetic field or by attaching a ferromagnet to a thin-plate chiral magnet. The created skyrmions are expected to behave as spin-torque oscillators in a series connection. Second, more intense readout signals can be expected for the skyrmion spin-wave reservoirs because the skyrmions directly interact via magnetic exchange interactions and thus are strongly coupled with each other as compared with the series-connected spin-torque oscillators interacting via weak magnetic dipole-dipole interactions or via indirect electric-circuit connections~\cite{Kanao2019,Furuta2018}.

In fact, a concept of spintronics reservoir using magnetic skyrmions was first proposed in Refs.~\cite{Prychynenko2018,Bourianoff2018,Pinna2020}, which is based on the electric-current injections to a skyrmion-hosting metallic magnet, that is, the electric currents are injected as input data, while variations of electric resistance due to skyrmions are measured as readout signals. On the contrary, our skyrmion reservoir is based on the spin-wave propagations in the skyrmion-hosting magnet. We expect that our skyrmion spin-wave reservoir can host remarkable nonlinearity because of complicated interferences of spin waves. A pronounced short-term memory effect can also be expected for our reservoir because of the slow damping of magnetization dynamics. Furthermore, we expect durability of the reservoir system because spin waves do not drive magnetic skyrmions so much and thus hardly affect their spatial positions~\cite{Mochizuki2012}, which might be of practical importance to achieve stable computations.
%%As will be argued later, our skyrmion spin-wave reservoir turned out to have better characteristics in terms of the nonlinearity and the short-term memory, both of which are figures-of-merit for the performance of reservoir. We expect that the spin-wave propagations or the magnetization dynamics are better for the reservoir application than the electric resistance. Namely, remarkable nonlinearity can be expected because of enhanced interference effects of the spin waves, which cannot be expected in electric currents. A pronounced short-term memory effect can also be expected because of the slow damping of magnetization dynamics. Furthermore, we expect durability of the reservoir system for our skyrmion spin-wave reservoir. In the previously proposed skyrmion reservoir based on the electric-resistance measurement, electric currents are injected into the reservoir system itself, which might drive the skyrmions through spin-transfer torques and cause displacements of the skyrmion positions. On the contrary, spin waves hardly affect the spatial positions of skyrmions, which might be of practical importance to achieve stable computations.

\subsection{Reservoir computing}
We examine potentials of magnetic skyrmions for application to reservoir computing by investigating several characteristics required for physical reservoirs, i.e., (1) the generalization ability~\cite{Nakane2018}, (2) the short-term memory~\cite{Kanao2019,Furuta2018}, and (3) the nonlinearity~\cite{Kanao2019,Furuta2018} in response. In our skyrmion reservoir, these properties are carried by spin waves, i.e., precession dynamics of magnetizations, propagating in a thin-plate magnet hosting skyrmions. In the present study, we deal with magnetic skyrmions packed in a rectangular-shaped thin-plate magnet, which form a distorted hexagonal lattice called skyrmion crystal. It is known that magnetic skyrmions can appear also in a random form depending on the magnetic-field strength or the sample quality. We expect that such a random configuration of skyrmions, which is referred to as skyrmion fabrics in Refs.~\cite{Prychynenko2018,Bourianoff2018,Pinna2020}, also exhibit similar reservoir functions.

A general procedure of computing with the skyrmion spin-wave reservoir is as follows. We take a set of input data for training, $\{s_{\rm in}^{\rm train}(k)\}$ ($k=1,2, \cdots, L_{\rm train}$), which can be either Boolean-type binary digits or continuous variables. Each data is entered to a skyrmion-hosting magnetic sample (reservoir) by an out-of-plane magnetic field pulse to trigger dynamics of magnetizations constituting the skyrmions. The induced magnetization dynamics propagate recursively within the reservoir, and after experiencing significant interferences, they finally reach the readout nodes (detectors). Then, the magnetization dynamics at the $N$ detectors are measured as components of the $N$-dimensional reservoir-state vector $\bm x(k)$, where
%%%%%%%%%%%%%%%%%%%%%%%%%%%%%%%%%%%%%%%%%%%%%%%%%%%%%%%%%%%%%
\begin{align}
\bm x(k)=
\begin{pmatrix}
x_1(k)\\
x_2(k)\\
\vdots \\
x_N(k)\\
\end{pmatrix}.
%%\left(x_1(k), x_2(k), \cdots, x_N(k)\right)^{\rm t}.
\end{align}
%%%%%%%%%%%%%%%%%%%%%%%%%%%%%%%%%%%%%%%%%%%%%%%%%%%%%%%%%%%%%
The component $x_n(k)$ represents a signal measured at the $n$th readout node by the $k$th measurement. The signals meaured by the $k$th measurement are translated to the $k$th output $y_{\rm out}(k)$ by a linear transformation using the $N$-dimensional weight vector ${\bm W}_{\rm out}$ as,
%%%%%%%%%%%%%%%%%%%%%%%%%%%%%%%%%%%%%%%%%%%%%%%%%%%%%%%%%%%%%
\begin{align}
y_{\rm out}(k)={\bm W}_{\rm out} \cdot \bm x(k).
\end{align}
%%%%%%%%%%%%%%%%%%%%%%%%%%%%%%%%%%%%%%%%%%%%%%%%%%%%%%%%%%%%%
The components of ${\bm W}_{\rm out}$ are $k$-independent and are optimized so as to output desired or correct values of $y_{\rm out}(k)$ for a given problem or task. This optimization is called training or learning. More specifically, we optimize the weight vector ${\bm W}_{\rm out}$ by minimizing the mean squared error (MSE) between the outputs $y_{\rm out}(k)$ and values desired or expected to be correct for the problem called targets $y_{\rm target}(k)$. Note that this translation is based on a linear transformation, and $y_{\rm out}(k)$ is given by a linear combination of $x_n(k)$ $(n=1,2, \cdots, N)$, while the nonlinear input-output transformation is thoroughly carried by the reservoir.

The MSE is given by,
%%%%%%%%%%%%%%%%%%%%%%%%%%%%%%%%%%%%%%%%%%%%%%%%%%%%%%%%%%%%%
\begin{eqnarray}
\text{MSE}
&=&\frac{1}{L_{\rm train}}\sum^{L_{\rm train}}_{k=1}
[y_{\rm target}(k)-y_{\rm out}(k)]^2
\nonumber\\
&=&\frac{1}{L_{\rm train}}\sum^{L_{\rm train}}_{k=1}
\left[y_{\rm target}(k)-{\bm W}_{\rm out} \cdot {\bm x}(k)
\right]^2.
\label{MSE}
\end{eqnarray}
%%%%%%%%%%%%%%%%%%%%%%%%%%%%%%%%%%%%%%%%%%%%%%%%%%%%%%%%%%%%%
The optimized weight vector ${\bm W}^{\rm opt}_{\rm out}$ that minimizes the MSE can be obtained using the pseudoinverse-matrix method~\cite{Fujii2017,Strang1993} as,
%%%%%%%%%%%%%%%%%%%%%%%%%%%%%%%%%%%%%%%%%%%%%%%%%%%%%%%%%%%%%
\begin{eqnarray}
{\bm W}^{\rm opt}_{\rm out}={\bm Y}^{T}_{\rm target} \hat{\bm X}^+.
\label{inverse}
\end{eqnarray}
%%%%%%%%%%%%%%%%%%%%%%%%%%%%%%%%%%%%%%%%%%%%%%%%%%%%%%%%%%%%%
Here ${\bm Y}_{\rm target}$ is the $L_{\rm train}$-dimensional vector composed of $y_{\rm target}(k)$, and $\hat{\bm X}^+$ is the $L_{\rm train} \times N$ pseudoinverse matrix of the $N\times L_{\rm train}$ matrix $\hat{\bm X}$ composed of vectors $\bm x_n(k)$~\cite{Fujii2017,Strang1993}. They are respectively given by,
%%%%%%%%%%%%%%%%%%%%%%%%%%%%%%%%%%%%%%%%%%%%%%%%%%%%%%%%%%%%%
\begin{eqnarray}
{\bm Y}_{\rm target}=
\begin{pmatrix}
y_{\rm target}(1)\\
y_{\rm target}(2)\\
\vdots \\
y_{\rm target}(L_{\rm train})\\
\end{pmatrix},
\end{eqnarray}
%%%%%%%%%%%%%%%%%%%%%%%%%%%%%%%%%%%%%%%%%%%%%%%%%%%%%%%%%%%%%
%%%%%%%%%%%%%%%%%%%%%%%%%%%%%%%%%%%%%%%%%%%%%%%%%%%%%%%%%%%%%
\begin{eqnarray}
\hat{\bm X}=
\begin{pmatrix}
x_1(1) & x_1(2) & \cdots & x_1(L_{\rm train}) \\
x_2(1) & x_2(2) & \cdots & x_2(L_{\rm train}) \\
\vdots & \vdots &        & \vdots \\
x_N(1) & x_N(2) & \cdots & x_N(L_{\rm train}) \\
\end{pmatrix}.
\end{eqnarray}
%%%%%%%%%%%%%%%%%%%%%%%%%%%%%%%%%%%%%%%%%%%%%%%%%%%%%%%%%%%%%

After the optimization of $\bm W_{\rm out}$ by sufficient training, we input another set of data called testing data $\{s_{\rm in}^{\rm test}(\ell)\}$ ($\ell=1,2, \cdots, L_{\rm test}$) into the skyrmion spin-wave reservoir. We again measure induced response signals at $N$ readout nodes and construct the $N$-dimensional reservoir-state vector $\bm x(\ell)$ for the $\ell$th measurement. We then translate it to the output $y_{\rm out}(\ell)$ using the optimized weight vector ${\bm W}^{\rm opt}_{\rm out}$, and compare thus obtained outputs $y_{\rm out}(\ell)$ with targets $y_{\rm target}(\ell)$ to check whether the reservoir can provide correct answers and/or desired outputs.

To investigate three of the required properties of reservoir, i.e., the generalization ability, the short-term memory function, and the nonlinearity, we impose the duration-estimation task, the short-term memory task, and the parity-check task on our skyrmion spin-wave reservoir, respectively. For each task, we choose appropriate magnetic-field pulses to represent the input data $\{s_{\rm in}^{\rm train}(k)\}$ and $\{s_{\rm in}^{\rm test}(\ell)\}$. We also properly define the reservoir-state vectors $\bm x(k)$ and the targets $y_{\rm target}(k)$ as argued in the following sections.

\subsection{Skyrmion crystal}
%%%%%%%%%%%%%%%%%%%%%%%%%%%%%%%%%%%%%%%%%%%%%%%%%%%%%%%%%%%%%
\begin{figure}[tb]
\centering
\includegraphics[scale=1.0]{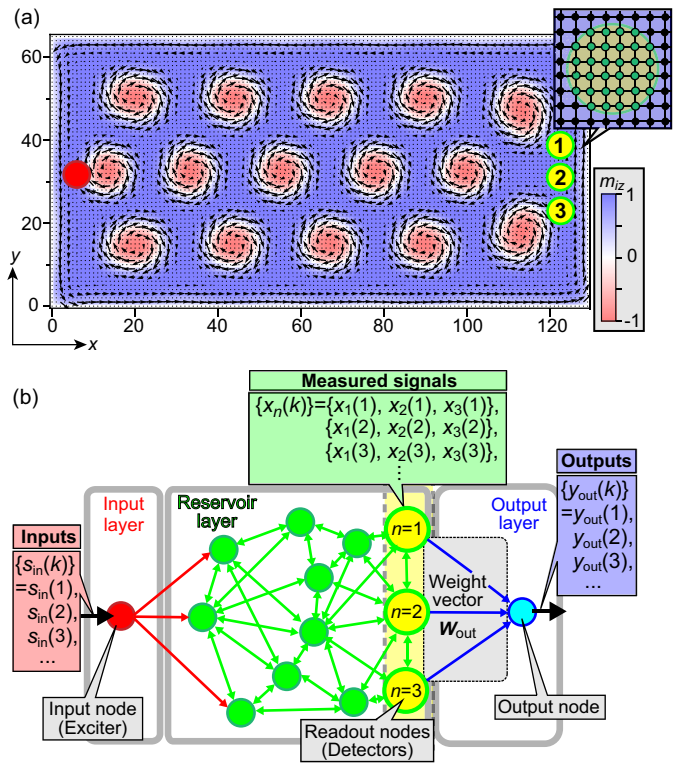}
\caption{(a) Magnetization configuration of the skyrmion spin-wave reservoir. Magnetic skyrmions are packed in a rectangular system of $128 \times 64$ sites described by the classical Heisenberg model on a square lattice in Eq.~(\ref{Hamiltonian}). In-plane magnetization vectors are presented by arrows at sites ($i_x$, $i_y$) when the integer coordinates $i_x$ and $i_y$ satisfy conditions mod($i_x$, 2)=1 and mod($i_y$, 2)=1, while the out-of-plane components are presented by colors. A setup of the input and readout nodes used for the duration-estimation task is also shown. The red circle near the left edge indicates the input area, and three green circles labeled by numbers 1, 2, and 3 near the right edge indicate the readout nodes (detectors), each of which has a radius of 3 in units of the lattice constant and contains 32 sites inside the area. Inset shows a magnified view of a detector. (b) Procedure of the reservoir computing for the duration-estimation task.}
\label{Fig02}
\end{figure}
%%%%%%%%%%%%%%%%%%%%%%%%%%%%%%%%%%%%%%%%%%%%%%%%%%%%%%%%%%%%%
To describe a skyrmion crystal in the skyrmion spin-wave reservoir, we employ a classical Heisenberg model on a square lattice. The Hamiltonian contains the nearest-neighbor ferromagnetic exchange interactions, the Zeeman interactions, and the Dzyaloshinskii-Moriya interactions (DMI) as,
%%%%%%%%%%%%%%%%%%%%%%%%%%%%%%%%%%%%%%%%%%%%%%%%%%%%%%%%%%%%
\begin{eqnarray}
\mathcal{H}
&=&-J\sum_{\langle i,j\rangle}\bm m_i \cdot \bm m_j
-\sum_i [\bm H+\bm H^\omega(\bm r_i,t)] \cdot \bm m_i
\nonumber\\
&+&D\sum_i (\bm m_i\times \bm m_{i+\hat{x}} \cdot \hat{x} 
+ \bm m_i\times \bm m_{i+\hat{y}} \cdot \hat{y}),
\label{Hamiltonian}
\end{eqnarray}
%%%%%%%%%%%%%%%%%%%%%%%%%%%%%%%%%%%%%%%%%%%%%%%%%%%%%%%%%%%%
where $\bm m_i$ is a classical magnetization vector at site $i$, length of which is normalized to be unity ($|\bm m_i|=1$). For the Zeeman-interaction term, we consider two kinds of magnetic fields, i.e., a DC magnetic field $\bm H$ and a time-dependent AC magnetic field $\bm H^\omega(\bm r_i,t)$,
%%%%%%%%%%%%%%%%%%%%%%%%%%%%%%%%%%%%%%%%%%%%%%%%%%%%%%%%%%%%
\begin{equation}
\bm H=H_z \hat{\bm z}, \quad \bm H^\omega(\bm r_i,t)=H^\omega(\bm r_i)\sin(\omega t) \hat{\bm z}.
\end{equation}
%%%%%%%%%%%%%%%%%%%%%%%%%%%%%%%%%%%%%%%%%%%%%%%%%%%%%%%%%%%%
The DC magnetic field $\bm H$ is applied globally to the entire system constantly, whereas the AC magnetic field $\bm H^\omega(\bm r_i,t)$ is applied locally to a restricted area regarded as an input node (exciter). For a practical device of the input node, we assume, for example, a micrometric metallic ring to generate a time-dependent local magnetic field within the ring via injecting a temporally varying electric current [see Fig.~\ref{Fig01}(b)]. Both $\bm H$ and $\bm H^\omega(\bm r_i,t)$ fields are applied perpendicular to the thin-plate plane of skyrmion-hosting magnet. 

%%%%%%%%%%%%%%%%%%%%%%%%%%%%%%%%%%%%%%%%%%%%%%%%%%%%%%%%%%%%
\begin{table}[tb]
\centering
\caption{Unit conversion table when $J=1\;{\rm meV}$.}
\begin{tabular}{lll} \hline \hline
\ & \begin{tabular}{c} Dimensionless \\ quantity \end{tabular}
& \begin{tabular}{c} Corresponding value \\ with units \end{tabular} \\ \hline
Exchange int. & $J=1$ & $1\;{\rm meV}$ \\
Time & $t=1000$ & $10^3 \times \hbar/J=0.66\;{\rm ns}$ \\
Frequency & $\omega=0.01$ & $10^{-2} \times J/2\pi\hbar=2.41\;{\rm GHz}$ \\
Magnetic field & $H=0.001$ & $10^{-3}\times J/\hbar\gamma=8.64\;{\rm mT}$ \\
\hline \hline
\end{tabular}
\label{tab:units}
\end{table}
%%%%%%%%%%%%%%%%%%%%%%%%%%%%%%%%%%%%%%%%%%%%%%%%%%%%%%%%%%%%
The position vector $\bm r_i=(i_x,i_y)$ represents the integer coordinates of site $i$ in units of the lattice constant. We take $J(\equiv 1)$ as energy units and choose $D=0.36$ for the strength of DMI. According to the saddle-point equation of the total energy, this DMI parameter leads to a skyrmion diameter of $2\pi/[\tan^{-1}(D/\sqrt{2}J)]\sim25$ in units of the lattice constant. In the following calculations, strength of the DC magnetic field is fixed at $H_z=0.06$, while the amplitude and frequency of the AC magnetic field are fixed at $H^\omega(\bm r)=0.008$ and $\omega=0.12416$, respectively. Note that this DC magnetic field leads to a stable skyrmion crystal in the ground state, while this AC magnetic field efficiently excites a spin-wave mode of skyrmion crystal called breathing mode when $D=0.36$, which are deduced from numerical calculation results for $D=0.09$ in Ref.~\cite{Mochizuki2012} after the following scale transformations with $a$=4~\cite{Schutte2014},
%%%%%%%%%%%%%%%%%%%%%%%%%%%%%%%%%%%%%%%%%%%%%%%%%%%%%%%%%%%%
\begin{equation}
D \rightarrow a D \;\Rightarrow\; 
|\bm H| \rightarrow a^2|\bm H|, \;\omega \rightarrow a^2 \omega.
\end{equation}
%%%%%%%%%%%%%%%%%%%%%%%%%%%%%%%%%%%%%%%%%%%%%%%%%%%%%%%%%%%%
The unit conversions from natural units to SI units are summarized in Table~\ref{tab:units}, where $\gamma \equiv g\mu_{\rm B}/\hbar$ is the gyromagnetic ratio.

The magnetization dynamics induced by applied AC magnetic fields are simulated by numerically solving the Landau-Lifshitz-Gilbert (LLG) equation using the fourth-order Runge-Kutta method. The equation is given by,
%%%%%%%%%%%%%%%%%%%%%%%%%%%%%%%%%%%%%%%%%%%%%%%%%%%%%%%%%%%%
\begin{eqnarray}
\frac{d\bm m_i}{dt}
=-\frac{1}{1+\alpha_{\rm G}^2}\left[
\bm m_i\times \bm H^{\rm eff}_i
+\alpha_{\rm G}\bm m_i \times (\bm m_i \times \bm H^{\rm eff}_i)
\right],\nonumber\\
\end{eqnarray}
%%%%%%%%%%%%%%%%%%%%%%%%%%%%%%%%%%%%%%%%%%%%%%%%%%%%%%%%%%%%
where $\alpha_{\rm G}(=0.1)$ is the dimensionless Gilbert-damping constant. The effective local magnetic fields $\bm H^{\rm eff}_i$ acting on the magnetization $\bm m_i$ are calculated by the $\bm m_i$-derivative of the Hamiltonian as,
%%%%%%%%%%%%%%%%%%%%%%%%%%%%%%%%%%%%%%%%%%%%%%%%%%%%%%%%%%%%
\begin{eqnarray}
\bm H^{\rm eff}_i = -\frac{\partial\mathcal{H}}{\partial \bm m_i}.
\end{eqnarray}
%%%%%%%%%%%%%%%%%%%%%%%%%%%%%%%%%%%%%%%%%%%%%%%%%%%%%%%%%%%%

First, we prepare an initial magnetization configuration in the absence of AC magnetic field $\bm H^\omega(\bm r_i, t)$ by setting $H^\omega(\bm r_i)=0$ in the Hamiltonian $\mathcal{H}$. Figure~\ref{Fig02}(a) presents the magnetization configuration of a distorted skyrmion crystal confined in a rectangular-shaped square-lattice system of 128 $\times$ 64 sites with open boundary conditions. This magnetization configuration is obtained by the Monte Carlo thermalization with simulated annealing to low temperatures, followed by a further relaxation under a sufficient time evolution in the LLG equation. Taking this configuration as an initial state of the reservoir, we apply local AC magnetic fields as inputs for the reservoir computing. Specifically, the AC magnetic field $\bm H^\omega(\bm r_i, t)$ is applied to a small circular area designed as an input node shown as a red circle in Fig.~\ref{Fig02}(a). The radius of the circle is 3, and its center is located at $(x,y)=(6.5, 32.5)$, both in units of the lattice constant.

\section{Results}
\subsection{Duration-estimation task}
%%%%%%%%%%%%%%%%%%%%%%%%%%%%%%%%%%%%%%%%%%%%%%%%%%%%%%%%%
\begin{figure*}[tb]
\centering
\includegraphics[scale=1]{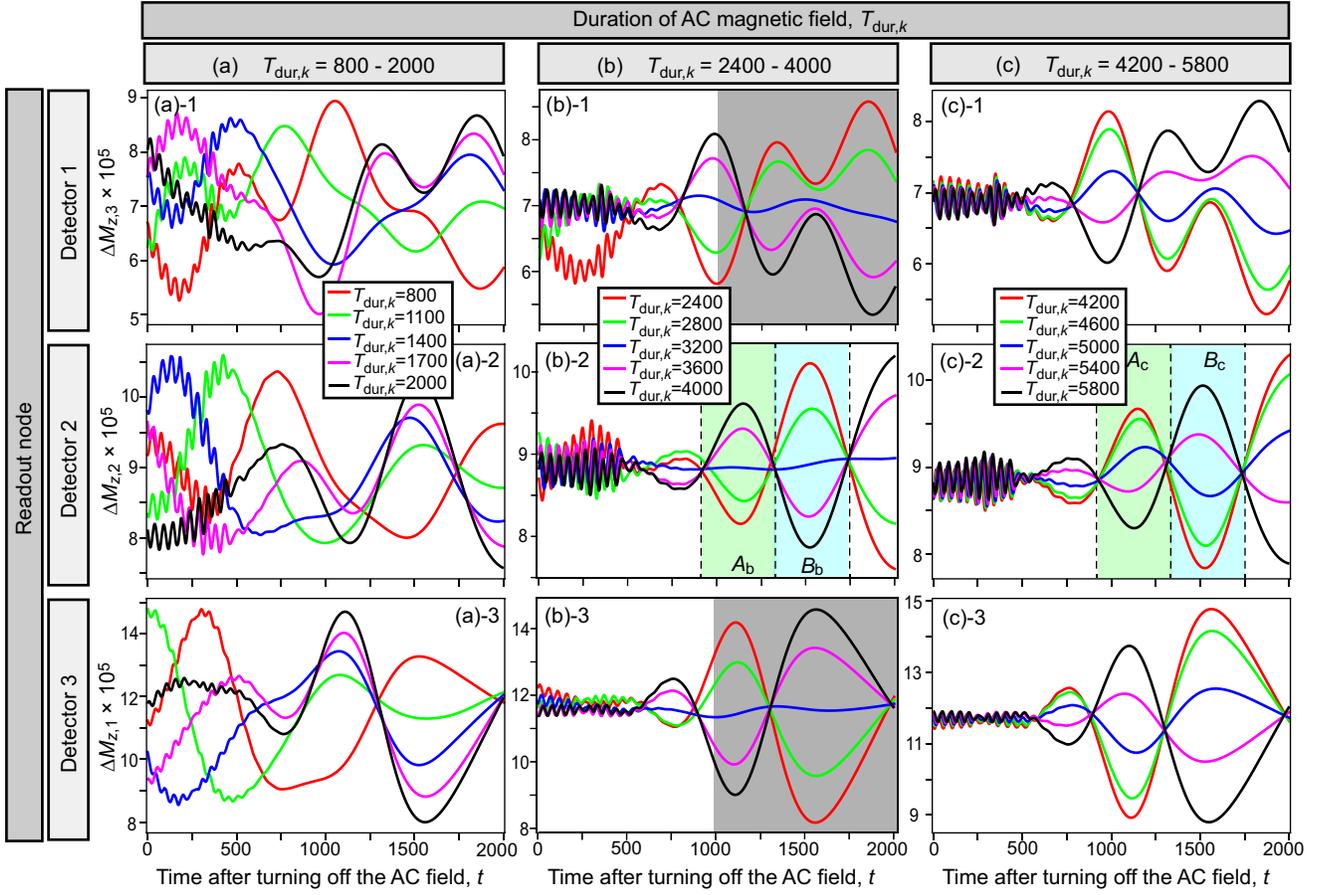}
\caption{Selected time profiles of the site-averaged magnetizations $\Delta M_{z,n}(T_{{\rm dur},k},t)$ at each detector after turning off the AC magnetic field with duration of $T_{{\rm dur},k}$ for various ranges of $T_{{\rm dur},k}$, i.e., (a) $T_{{\rm dur},k}$=800-2000, (b) $T_{{\rm dur},k}$=2400-4000, and (c) $T_{{\rm dur},k}$=4200-5800, where $t$ is time after turning off the AC magnetic field. In each duration range, three different detectors labeled by $n$=1, 2, and 3 shown in Fig.~\ref{Fig02}(a) are examined. We define $\Delta M_{z,n}(T_{{\rm dur},k},t)$ as $\Delta M_{z,n}(T_{{\rm dur},k},t) \equiv M_{z,n}(T_{{\rm dur},k},t)-M_{{\rm s},n}$ where $M_{z,n}(T_{{\rm dur},k},t)=\sum_i m_{z,i}(T_{{\rm dur},k},t)/N_{\rm detect}$ is the out-of-plane component of magnetization averaged over the sites within the area of detector, and $M_{{\rm s},n}$ is a steady component. Here $N_{\rm detect}$(=32) is the number of sites within the area of each detector, while the values of $M_{{\rm s},n}$ are chosen to be $M_{{\rm s},n}=0.9467$ for Detectors 1 and 3, whereas $M_{{\rm s},n}=0.9412$ for Detector 2. Shaded areas in (b)-1 and (b)-3 indicate time ranges used to calculate the temporal averages of magnetization in the following Fig.~\ref{Fig04}(a) for the duration-estimation task.}
\label{Fig03}
\end{figure*}
%%%%%%%%%%%%%%%%%%%%%%%%%%%%%%%%%%%%%%%%%%%%%%%%%%%%%%%%%
We first investigate the duration-estimation task. By imposing this task, we examine whether the skyrmion spin-wave reservoir can correctly estimate the durations of input AC magnetic fields as a test for the ability to evaluate unknown variables of input signals. For this task, various durations $T_{{\rm dur},k}$ of the AC magnetic field are chosen as inputs $s_{\rm in}(k)$ [see also Fig.~\ref{Fig02}(b)], 
%%%%%%%%%%%%%%%%%%%%%%%%%%%%%%%%%%%%%%%%%%%%%%%%%%%%%%%%%
\begin{align}
s_{\rm in}(k)=T_{{\rm dur},k}.
\end{align}
%%%%%%%%%%%%%%%%%%%%%%%%%%%%%%%%%%%%%%%%%%%%%%%%%%%%%%%%%
We apply a pulse of AC magnetic field during a time range of $0 \leq \bar{t} < T_{{\rm dur},k}$ to a small circular area designed as an input node, and, subsequently, turn it off at $\bar{t}=T_{{\rm dur},k}$ and let the magnetizations relax. We trace the magnetization dynamics during this process by numerically solving the LLG equation.

The detectors are three circular (green) areas labeled by numbers $n$=1, 2, and 3 in Fig.~\ref{Fig02}(a). Their radii are 3 in units of the lattice constant, and their centers are located at $(x, y)=(122.5, 40.5)$, $(122.5, 32.5)$ and $(122.5, 24.5)$, respectively. We define $t$ as time after turning off the AC magnetic field (i.e., $t \equiv \bar{t}-T_{{\rm dur},k}$). At each detector, we measure dynamics of local magnetizations. The time profile of $z$-component magnetization averaged over sites contained in the area of $n$th detector is given by,
%%%%%%%%%%%%%%%%%%%%%%%%%%%%%%%%%%%%%%%%%%%%%%%%%%%%%%%%%
\begin{align}
M_{z,n}(T_{{\rm dur},k}, t)
=\frac{1}{N_{\rm detect}}
\sum_{i\in n\text{th detector}} m_{z,i}(T_{{\rm dur},k}, t),
\end{align}
%%%%%%%%%%%%%%%%%%%%%%%%%%%%%%%%%%%%%%%%%%%%%%%%%%%%%%%%%
where $N_{\rm detect}(=32)$ is the number of lattice sites within each detector area. We further subtract a steady component $M_{{\rm s},n}$ as,
%%%%%%%%%%%%%%%%%%%%%%%%%%%%%%%%%%%%%%%%%%%%%%%%%%%%%%%%%
\begin{align}
\Delta M_{z,n}(T_{{\rm dur},k}, t)=M_{z,n}(T_{{\rm dur},k}, t) - M_{{\rm s},n},
\end{align}
%%%%%%%%%%%%%%%%%%%%%%%%%%%%%%%%%%%%%%%%%%%%%%%%%%%%%%%%%
where $M_{{\rm s},1}=M_{{\rm s},3}=0.9467$ for Detectors 1 and 3, and $M_{{\rm s},2}=0.9412$ for Detector 2. 

Figure~\ref{Fig03} shows simulated time profiles of $\Delta M_{z,n}(T_{{\rm dur},k}, t)$ for three different ranges of durations $T_{{\rm dur},k}$, i.e., (a) shorter durations ($T_{{\rm dur},k}$=800-2000), (b) intermediate durations ($T_{{\rm dur},k}$=2400-4000), and (c) longer durations ($T_{{\rm dur},k}$=4200-5800) at Detectors 1, 2 and 3. The plots of $\Delta M_{z,n}(T_{{\rm dur},k}, t)$ exhibit temporal oscillations at respective detectors. In these plots, the behaviors of $\Delta M_{z,n}(T_{{\rm dur},k}, t)$ show clear dependence on the duration $T_{{\rm dur},k}$. We now focus on the plots of $\Delta M_{z,2}(T_{{\rm dur},k}, t)$ at Detector 2 for intermediate durations in Fig.~\ref{Fig03}(b)-2. We find that as $T_{\rm dur}$ increases, $\Delta M_{z,n}(T_{{\rm dur},k}, t)$ monotonically increases in the time range of $800 \leq t \leq 1300$ (Range $A_{\rm b}$), whereas it monotonically decreases in the subsequent time range of $1300 \leq t \leq 1750$ (Range $B_{\rm b}$). We also find that the amplitudes of oscillation monotonically increase for respective durations as time proceeds. Such behaviors are observed also in Detectors 1 and 3. These facts indicate that the time profiles of $\Delta M_{z,n}(T_{{\rm dur},k}, t)$ ($n$=1,2, and 3) involve information of the duration $T_{{\rm dur},k}$ of AC magnetic field.

%%%%%%%%%%%%%%%%%%%%%%%%%%%%%%%%%%%%%%%%%%%%%%%%%%%%%%%%%
\begin{figure}[tb]
\centering
\includegraphics[scale=1.0]{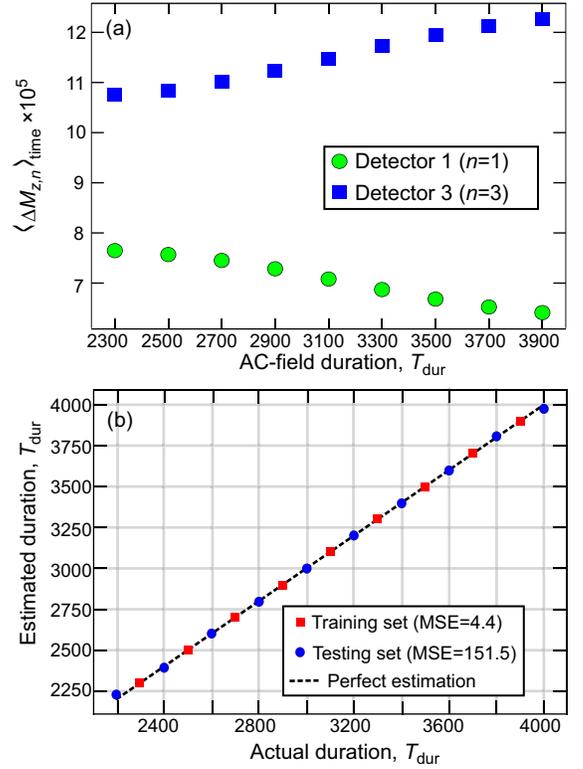}
\caption{(a) Temporal averages of dynamical magnetizations $\braket{\Delta M_{z,n}(T_{{\rm dur},k})}_{\rm time}$ at Detectors 1 and 3 ($n$=1 and 3) over time ranges indicated by shaded areas in Fig.~\ref{Fig03}(b)-1 and Fig.~\ref{Fig03}(b)-3 as functions of the duration of AC magnetic field for the training dataset. We adopt a polynomial form with respect to $\braket{\Delta M_{z,1}(T_{{\rm dur},k})}_{\rm time}$ and $\braket{\Delta M_{z,3}(T_{{\rm dur},k})}_{\rm time}$ in Eq.~(\ref{outputDET}) for the output $y_{\rm out}(k)$. (b) Results of the duration-estimation task. The horizontal (vertical) axis labels the actual (estimated) duration $T_{\rm dur}$ for both the training (red squares) and testing (blue circles) datasets. The dashed line with a slope of unity is the perfect estimation line for reference, indicating that amazingly accurate estimations are achieved.}
\label{Fig04}
\end{figure}
%%%%%%%%%%%%%%%%%%%%%%%%%%%%%%%%%%%%%%%%%%%%%%%%%%%%%%%%%
We adopt nine values of $T_{{\rm dur},k}$ in the intermediate duration range, i.e., $T_{{\rm dur},k}=2300+200k$ with $k=0, 1, 2, \cdots, 8$ as a set of input data for training. After applying the AC magnetic field for a certain duration of $T_{{\rm dur},k}$, we compute time-averages $\braket{\Delta M_{z,1}(T_{{\rm dur},k})}_{\rm time}$ and $\braket{\Delta M_{z,3}(T_{{\rm dur},k})}_{\rm time}$ at Detectors 1 and 3 over a time range of $1000 \leq t \leq 2000$ indicated by shaded areas in Fig.~\ref{Fig03}(b)-1 and Fig.~\ref{Fig03}(b)-3. We regard these time-averaged magnetizations as reservoir states. Figure~\ref{Fig04}(a) presents a plot of these quantities as functions of $T_{{\rm dur},k}$ for Detectors 1 and 3. This plot shows nearly linear behaviors with respect to $T_{{\rm dur},k}$ for both detectors, which stem from the monotonic trends of $\Delta M_{z,n}(T_{{\rm dur},k}, t)$ with respect to $T_{{\rm dur},k}$ as argued above.

Inspired by this characteristic, we devise the following polynomial form with respect to $\braket{\Delta M_{z,1}(T_{{\rm dur},k})}_{\rm time}$ and $\braket{\Delta M_{z,3}(T_{{\rm dur},k})}_{\rm time}$ for the output $y_{\rm out}(k)$,
%%%%%%%%%%%%%%%%%%%%%%%%%%%%%%%%%%%%%%%%%%%%%%%%%%%%%%%%%
\begin{equation}
y_{\rm out}(k)=W_0
+\sum^{m_{\rm max}}_{m=1} \left[
W_{2m-1} x_1^m(T_{{\rm dur},k}) + W_{2m} x_3^m(T_{{\rm dur},k})
\right],
\label{outputDET}
\end{equation}
%%%%%%%%%%%%%%%%%%%%%%%%%%%%%%%%%%%%%%%%%%%%%%%%%%%%%%%%%
where
%%%%%%%%%%%%%%%%%%%%%%%%%%%%%%%%%%%%%%%%%%%%%%%%%%%%%%%%%
\begin{equation}
x_n(T_{{\rm dur},k}) \equiv \braket{\Delta M_{z,n}(T_{{\rm dur},k})}_{\rm time} \times 10^5
\quad (n=1, 3).
\end{equation}
%%%%%%%%%%%%%%%%%%%%%%%%%%%%%%%%%%%%%%%%%%%%%%%%%%%%%%%%%
Here $W_0$ is a constant bias, and $m_{\rm max}(=3)$ is the maximal power of this polynomial model. The coefficients of linear combination, $W_{2m-1}$ and $W_{2m}$ ($m=1, 2, \cdots, m_{\rm max}$), are components of the weight vector ${\bm W}_{\rm out}$. In total, there are $2m_{\rm max}+1$ components in ${\bm W}_{\rm out}$.

Recalling that the duration-estimation task requires the reservoir to correctly estimate the durations of AC magnetic fields, we set the desired outputs, i.e., the targets $y_{\rm target}(k)$ to be the durations $T_{{\rm dur},k}$ themselves as,
%%%%%%%%%%%%%%%%%%%%%%%%%%%%%%%%%%%%%%%%%%%%%%%%%%%%%%%%%
\begin{align}
y_{\rm target}(k)=T_{{\rm dur},k}=s_{\rm in}(k).
\label{targetDET}
\end{align}
%%%%%%%%%%%%%%%%%%%%%%%%%%%%%%%%%%%%%%%%%%%%%%%%%%%%%%%%%
We substitute the outputs $y_{\rm out}(k)$ in Eq.~(\ref{outputDET}) and the targets $y_{\rm target}(k)$ in Eq.~(\ref{targetDET}) into Eq.~(\ref{MSE}), and then use Eq.~(\ref{inverse}) to obtain the optimal weight vector $\bm W^{\rm opt}_{\rm out}$, which minimize the MSE for the training dataset. After this training procedure, we plug $\bm W_{\rm out}^{\rm opt}$ into Eq.~(\ref{outputDET}) to estimate the durations $T_{{\rm dur},\ell}$ for ten testing data of $T_{{\rm dur},\ell}=2200+200\ell$ with $\ell=0, 1, \cdots, 9$ as well as the nine training data of $T_{{\rm dur},k}=2300+200k$ with $k=0, 1, \cdots, 8$.

In Fig.~\ref{Fig04}(b), we show a plot of the estimated durations (outputs) versus the actual durations (inputs). Here the dashed line with a slope of unity is the perfect-estimation line. The plot shows amazingly accurate estimations for both the training and testing datasets. Surprisingly, the MSE turns out to be as small as $151.5$ for the testing dataset. Considering that the square of the input values is in the order of $\sim 10^6$, we find that this MSE value is extremely small. We take the root mean square error (RMSE) divided by the input average as a dimensionless quantity to represent the accuracy of estimations. Since the average of durations for the testing dataset is 3100, this quantity is evaluated to be $\sqrt{151.5}/3100\sim 0.004$. This value is one order of magnitude smaller than that of another previously proposed spin-wave reservoir based on a ferromagnetic garnet film~\cite{Nakane2018}. In that work, it was assumed that the spin waves are excited by locally changing the uniaxial magnetic anisotropy and are exploited as reservoir states to estimate the durations of anisotropy change. According to Fig.~10 in Ref.~\cite{Nakane2018}, their RMSE divided by input average of this garnet-based spin-wave reservoir can be roughly estimated to be $\sim 0.08$. The present result demonstrates that our skyrmion spin-wave reservoir harbors a great generalization ability to estimate the unknown AC-field durations in the testing dataset and possesses high potential for application to machine-learning information processing.

Now we discuss a possible limitation of our skyrmion spin-wave reservoir in the generalization ability, that is, it might be difficult to correctly estimate widely ranging durations by our skyrmion system. More specifically, in order to correctly evaluate a certain duration, the weight vector $\bm W_{\rm out}$ should be trained by using a set of training data for the corresponding duration range. Namely, the weight vector $\bm W_{\rm out}$ trained by the data for intermediate durations can correctly estimate intermediate durations, but it might not be able to estimate longer or shorter durations. This is because the duration dependence of the behavior of magnetization dynamics varies depending on the duration range. To see this aspect, we compare the time profiles of $\Delta M_{z,2}(T_{\rm dur}, t)$ for the longer durations in Fig.~\ref{Fig03}(c)-2 and those for the intermediate durations in Fig.~\ref{Fig03}(b)-2. The magnetization $\Delta M_{z,2}(T_{\rm dur}, t)$ for the longer durations in Fig.~\ref{Fig03}(c)-2 monotonically decreases with increasing $T_{\rm dur}$ in the time range of $800 \leq t \leq 1300$ (Range $A_{\rm c}$), whereas it monotonically increases in the subsequent time range of $1300 \leq t \leq 1750$ (Range $B_{\rm c}$). These behaviors are opposite to the above-argued behaviors of $\Delta M_{z,2}(T_{\rm dur}, t)$ for the intermediate durations in Fig.~\ref{Fig03}(b)-2. Therefore, the weight vector $\bm W_{\rm out}$ optimized for the intermediate durations might not be able to estimate the shorter or longer durations correctly. In practical experiments, it might be required to presume or restrict the range of durations.

\subsection{Short-term memory task and parity-check task}
%%%%%%%%%%%%%%%%%%%%%%%%%%%%%%%%%%%%%%%%%%%%%%%%%%%%%%%%%
\begin{figure*}[tb]
\centering
\includegraphics[scale=1.0]{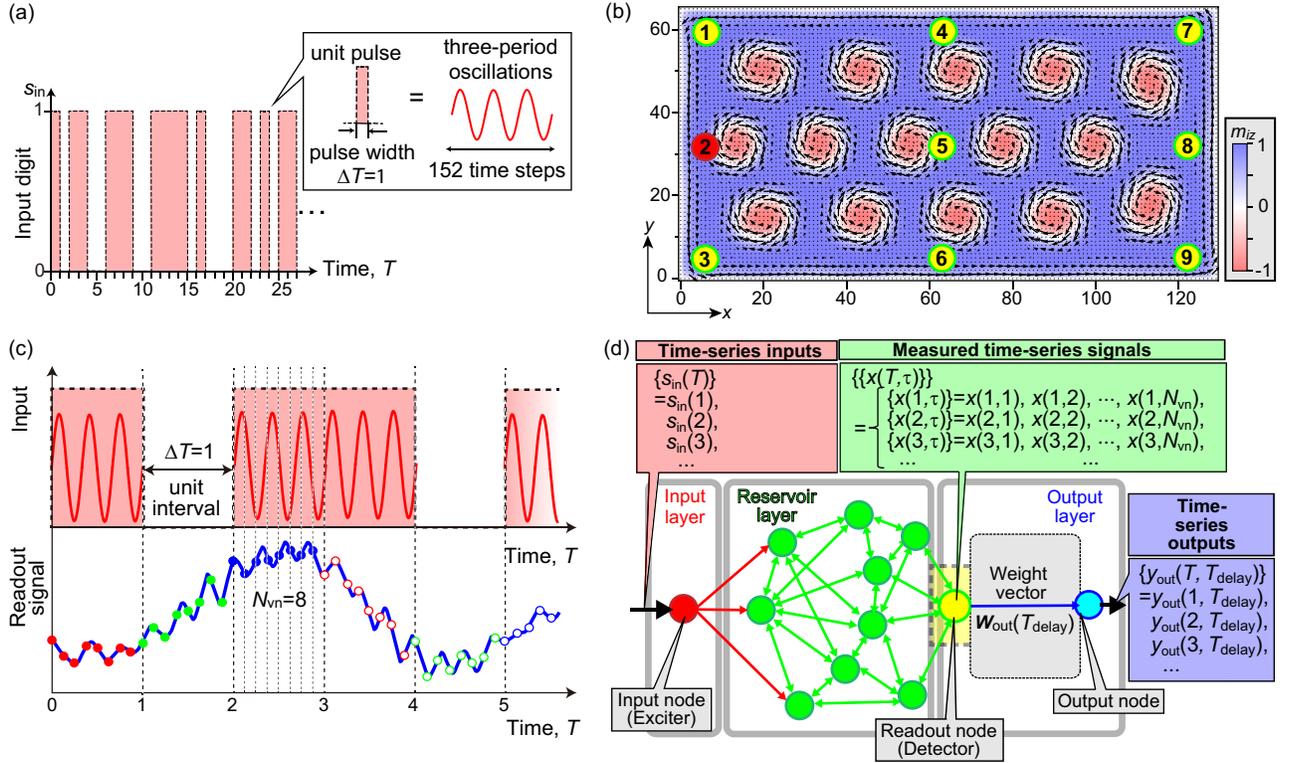}
\caption{(a) Schematics of the random time-sequence of binary digits ``1" and ``0" as input signals used for the STM and PC tasks. The input ``1" is represented by a pulse of three-period oscillations of AC magnetic field, while the input ``0" is represented by turning off the AC magnetic field for the same duration. The duration of pulse is 152 integer time steps for integration of the LLG equation, which is chosen as a unit of time. (b) Setup of the input node (red circle) and readout nodes (red circle and green circles) on the skyrmion spin-wave reservoir used for the STM and PC tasks. Each node area has a radius of 3 in units of the lattice constant and contains 32 sites inside it. For the readout nodes, nine different positions labeled by numbers 1 to 9 are examined. The red circle labeled by 2 is assigned to both input and readout nodes. (c) Concept of the virtual nodes. For each time $T$, the unit interval of $\Delta T=1$ from $T$ to $T+1$ that corresponds to the pulse width is divided into $N_{\rm vn}$(=8 in the present figure) moments with equivalent intervals. The signals are measured at $N_{\rm vn}$ moments at the readout node, which constitute the $(N_{\rm vn}+1)$-dimensional reservoir-state vector $\bm x(T)$. (d) Reservoir computing procedures for the STM and PC tasks.}
\label{Fig05}
\end{figure*}
%%%%%%%%%%%%%%%%%%%%%%%%%%%%%%%%%%%%%%%%%%%%%%%%%%%%%%%%%
Next we examine the short-term memory (STM) task and the parity-check (PC) task~\cite{Furuta2018,Kanao2019}. By imposing the STM task, we evaluate the short-term memory function of our skyrmion spin-wave reservoir, i.e., how long the reservoir can memorize the information of past input sequence. This property is crucially important to analyze time-series data for, e.g., market forecasts, sentence predictions, and voice/speech recognitions~\cite{Tanaka2019,Nakajima2020,Nakajima2018}. On the other hand, by imposing the PC task, we evaluate the ability of our reservoir to nonlinearly transform the input signals into readout signals. This property is indispensable to solve linearly inseparable problems~\cite{Tanaka2019,Nakajima2020,Nakajima2018,Bishop2006} through mapping the input data onto a higher-dimensional information space for, e.g., pattern classifications and hand-written digit recognitions~\cite{Bishop2006}.

For both tasks, the input data $s_{\rm in}(T)$ are time-series binary digits of ``1" and ``0", each of which is chosen randomly at every integer time $T$ [Fig.~\ref{Fig05}(a)]. When the input digit is ``1" at time $T$, we locally apply a pulse of three-period oscillations of AC magnetic field, duration of which corresponds to 152 integration time steps of the LLG equation. The magnetic field pulse is applied within a red circular area labeled by 2 in Fig.~\ref{Fig05}(b), center of which is located at $(x,y)=(6.5, 32.5)$. On the other hand, when the input digit is ``0", we turn off the AC magnetic field and let the magnetizations relax for the same duration of 152 integration time steps. Here the integer time $T$ is counted in units of this three-period duration.

We define the desired outputs $y_{\rm target}$ in Eq.~(\ref{MSE}) for respective tasks as,
%%%%%%%%%%%%%%%%%%%%%%%%%%%%%%%%%%%%%%%%%%%%%%%%%%%%%%%%%
\begin{align}
&y_{\rm target}^{\rm STM}(T,T_{\rm delay})=s_{\rm in}(T-T_{\rm delay}),
\label{STM}
\nonumber \\
\\
&y_{\rm target}^{\rm PC}(T,T_{\rm delay})={\rm mod}
\left[s_{\rm in}(T) + s_{\rm in}(T-1) \right. \nonumber\\
&\hspace*{3.5cm} \left. + \cdots +s_{\rm in}(T-T_{\rm delay}), 2\right],
\label{PC}
\end{align}
%%%%%%%%%%%%%%%%%%%%%%%%%%%%%%%%%%%%%%%%%%%%%%%%%%%%%%%%%
where $T_{\rm delay}$ is an integer variable that represents a given delay time. The STM task examines to what extent the input signal at a previous time $T-T_{\rm delay}$ can be reconstructed from the current reservoir states at time $T$. On the other hand, the PC task examines to what extent the reservoir can describe nonlinear functions by taking parity of the sum of past binary inputs from current time $T$ to a previous time $T-T_{\rm delay}$ as a typical example of nonlinear functions. Mapping of the input data onto a high-dimensional information space via nonlinear transformations is a key function of the reservoir, and it must be carried by the reservoir instead of the output layer because the output layer in reservoir computing simply produces outputs by a linear combination of the reservoir states~\cite{Tanaka2019,Nakajima2020,Nakajima2018}.

%%%%%%%%%%%%%%%%%%%%%%%%%%%%%%%%%%%%%%%%%%%%%%%%%%%%%%%%%%%%%
\begin{figure*}[tb]
\centering
\includegraphics[scale=1.0]{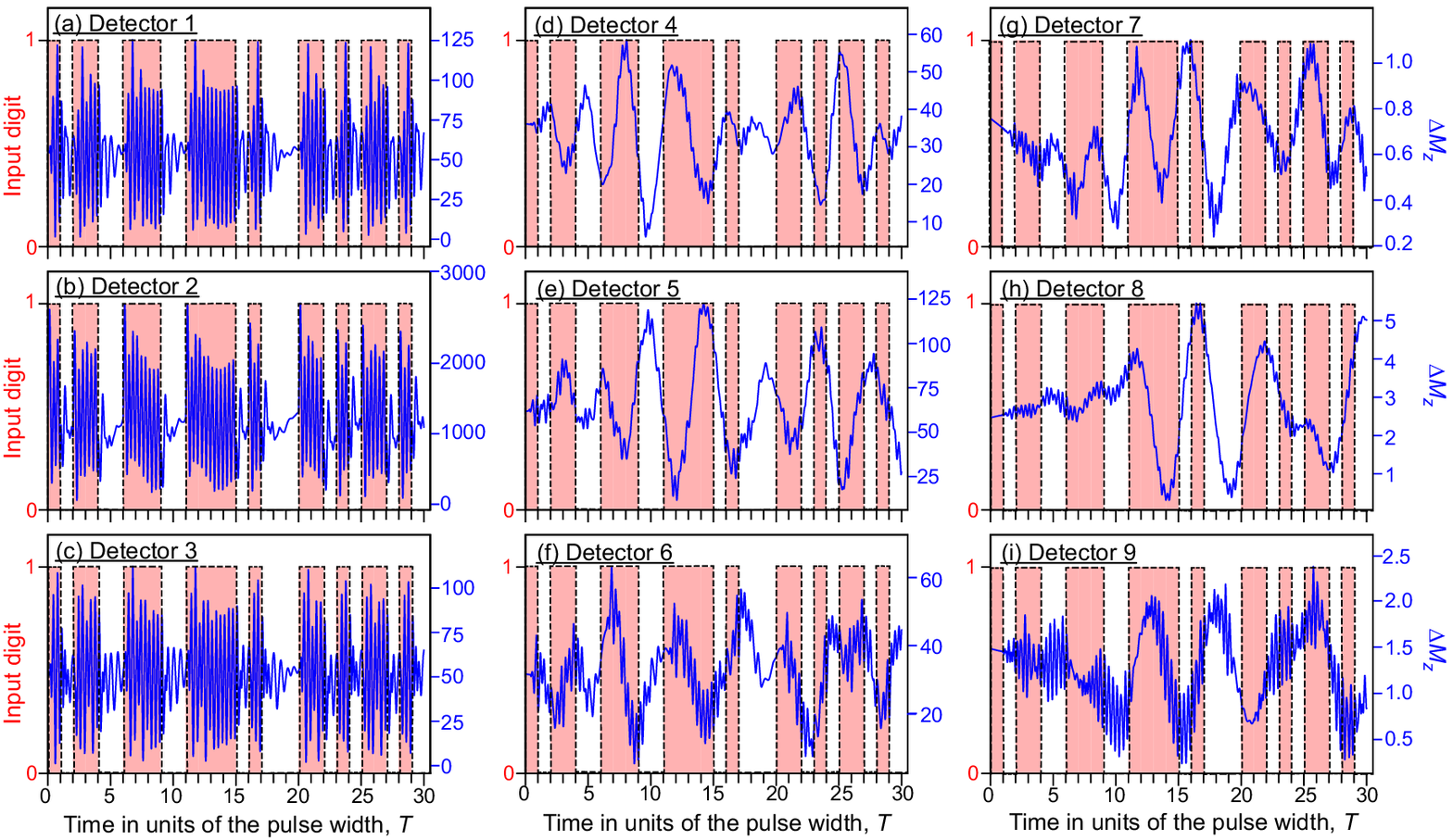}
\caption{(a)-(i) Time profiles of the dynamical magnetization $\Delta M_z(t)$ measured at Detectors 1-9 as responses to the first 30 random sequence of on/off AC-field pulses. A single AC-field pulse is composed of three-period oscillations of AC magnetic field, and its duration is chosen as a unit of time. The shaded rectangles and blanks in each panel indicate the time-series of input digits ``1" and ``0" entered to the reservoir as on/off of the AC-field pulses.}
\label{Fig06}
\end{figure*}
%%%%%%%%%%%%%%%%%%%%%%%%%%%%%%%%%%%%%%%%%%%%%%%%%%%%%%%%%%%%%
%%%%%%%%%%%%%%%%%%%%%%%%%%%%%%%%%%%%%%%%%%%%%%%%%%%%%%%%%%%%%
\begin{figure}[tb]
\centering
\includegraphics[scale=1.0]{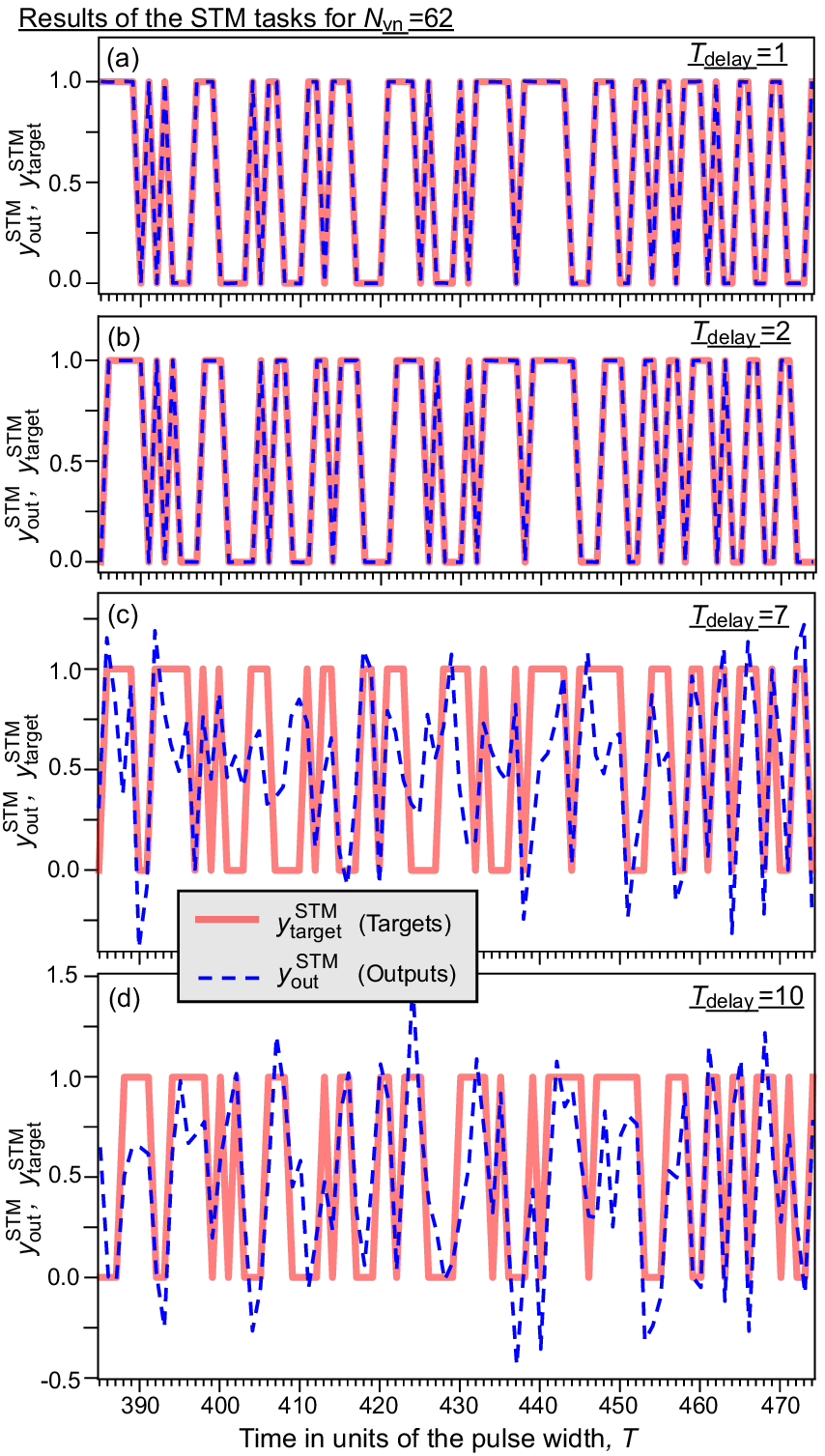}
\caption{Comparison between profiles of output $y_{\rm out}(T,T_{\rm delay})$ (blue dashed lines) and those of target $y_{\rm target}(T,T_{\rm delay})$ (red solid lines) for the STM task for various delays $T_{\rm delay}$, i.e., (a) $T_{\rm delay}$=1, (b) $T_{\rm delay}$=2, (c) $T_{\rm delay}$=7, and (d) $T_{\rm delay}$=10. The profiles of target are calculated by Eq.~(\ref{STM}), and the analyses are performed by setting the number of virtual nodes as $N_{\rm vn}$=62.}
\label{Fig07}
\end{figure}
%%%%%%%%%%%%%%%%%%%%%%%%%%%%%%%%%%%%%%%%%%%%%%%%%%%%%%%%%%%%%
%%%%%%%%%%%%%%%%%%%%%%%%%%%%%%%%%%%%%%%%%%%%%%%%%%%%%%%%%%%%%
\begin{figure}[tb]
\centering
\includegraphics[scale=1.0]{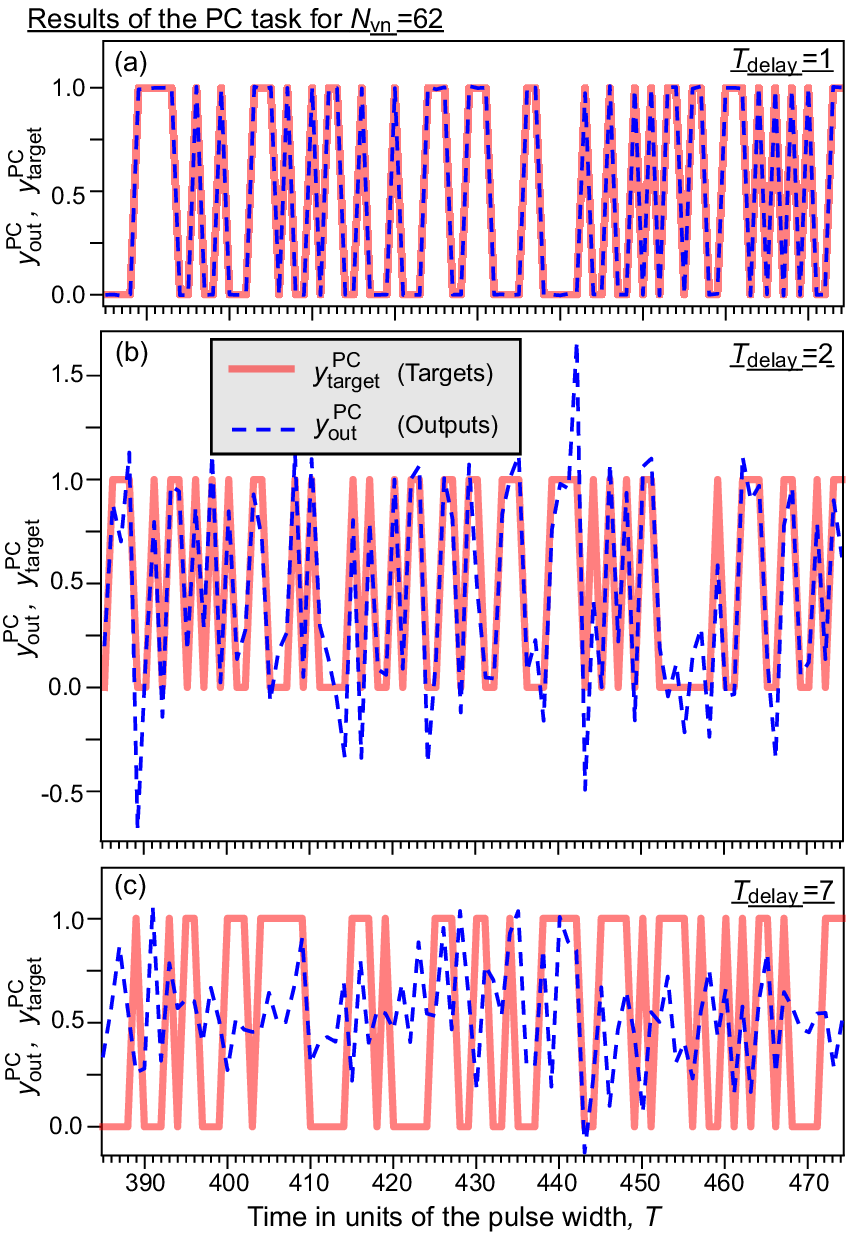}
\caption{Comparison between profiles of output $y_{\rm out}(T,T_{\rm delay})$ (blue dashed lines) and those of target $y_{\rm target}(T,T_{\rm delay})$ (red solid lines) for the PC task for various delays $T_{\rm delay}$, i.e., (a) $T_{\rm delay}$=1, (b) $T_{\rm delay}$=2, and (c) $T_{\rm delay}$=7. The profiles of target are calculated by Eq.~(\ref{PC}), and the analyses are performed by setting the number of virtual nodes as $N_{\rm vn}$=62.}
\label{Fig08}
\end{figure}
%%%%%%%%%%%%%%%%%%%%%%%%%%%%%%%%%%%%%%%%%%%%%%%%%%%%%%%%%%%%%
%%%%%%%%%%%%%%%%%%%%%%%%%%%%%%%%%%%%%%%%%%%%%%%%%%%%%%%%%%%%%
\begin{figure}[tb]
\centering
\includegraphics[scale=1.0]{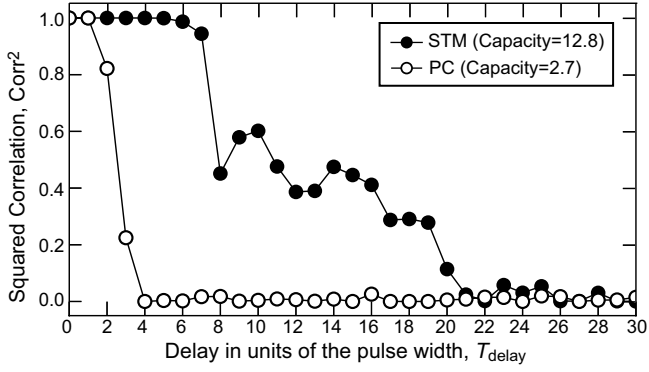}
\caption{Squared correlations, Corr$^2$, of the STM (filled circles) and PC (open circles) tasks as functions of the delay time $T_{\rm delay}$ for Detector 7. Capacities of the STM and PC tasks, $C_{\rm STM}$ and $C_{\rm PC}$, which correspond to areas below the respective curves, quantify the performances of reservoir for these tasks. The measurements of signals are performed by setting the number of virtual nodes as $N_{\rm vn}$=62.}
\label{Fig09}
\end{figure}
%%%%%%%%%%%%%%%%%%%%%%%%%%%%%%%%%%%%%%%%%%%%%%%%%%%%%%%%%%%%%
%%%%%%%%%%%%%%%%%%%%%%%%%%%%%%%%%%%%%%%%%%%%%%%%%%%%%%%%%%%%%
\begin{figure*}[tb]
\centering
\includegraphics[scale=1.0]{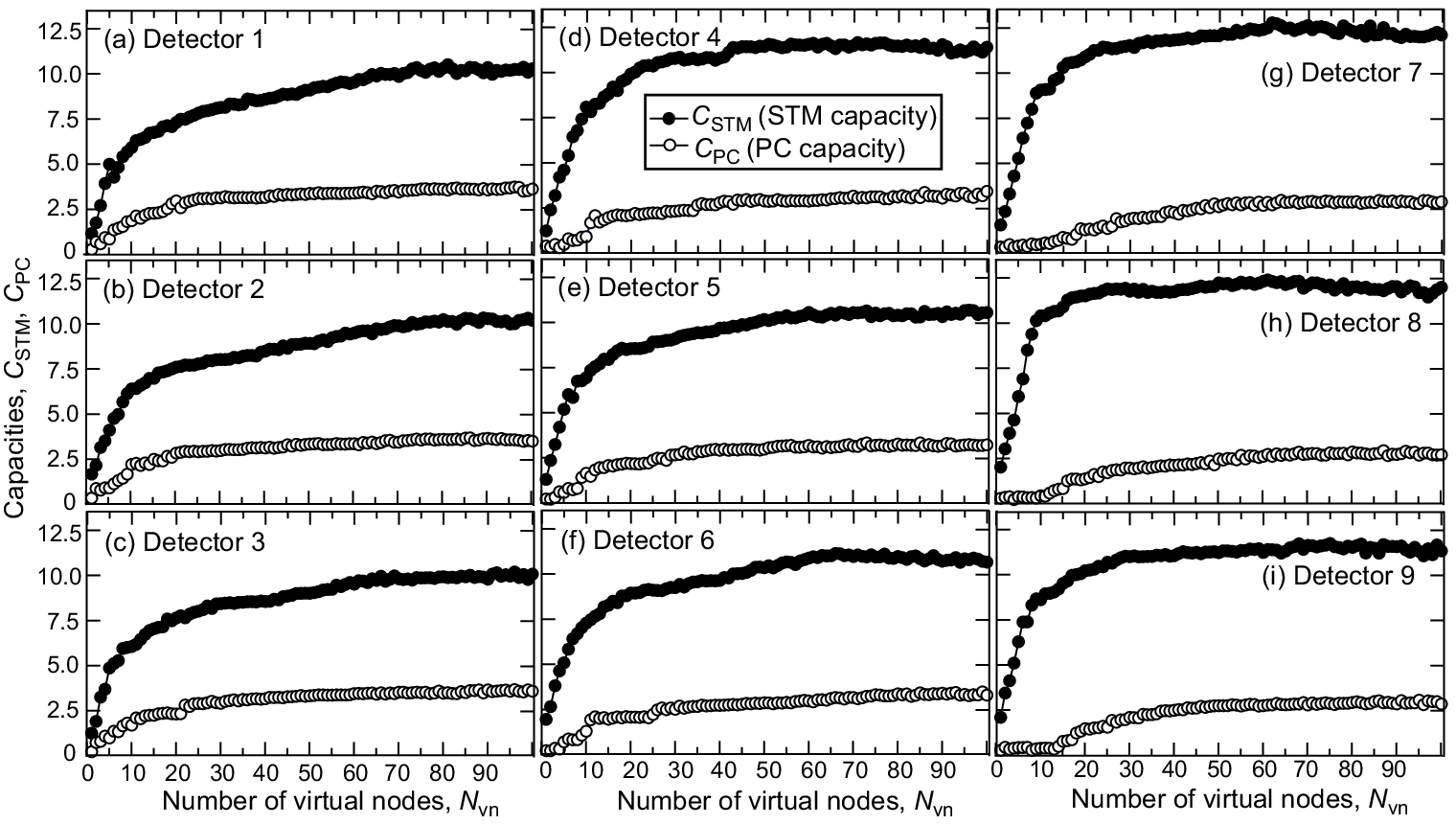}
\caption{Capacities of STM (filled circles) and PC (open circles) tasks, $C_{\rm STM}$ and $C_{\rm PC}$, as functions of the number of virtual nodes $N_{\rm vn}$ for respective detectors. The capacities correspond to summations of the squared correlations, Corr$^2$, over the range of $0 \leq T_{\rm delay} \leq 30$ for 90 testing data, after optimizing the weight matrix using 350 training data. For our skyrmion spin-wave reservoir, the largest values of $C_{\rm STM}$ and $C_{\rm PC}$ are 12.8 and 2.7, respectively.}
\label{Fig10}
\end{figure*}
%%%%%%%%%%%%%%%%%%%%%%%%%%%%%%%%%%%%%%%%%%%%%%%%%%%%%%%%%%%%%

Here we consider only one readout node for both tasks. In the reservoir computing, we need multiple readout data for a certain input to construct a reservoir-state vector. In the examination of the duration-estimation task, we have measured signals at multiple readout nodes, i.e., Detectors 1 and 3 [Fig.~\ref{Fig02}(a)] to prepare the multiple readout data. In the present examinations, instead of measuring signals at multiple readout nodes, we measure signals from a single readout node at multiple moments during the interval from $T$ to $T+1$, and regard the measured sequential data as the components of reservoir-state vector at time $T$. Specifically, we divide each unit interval of pulse $\Delta T$=1, i.e., a duration of the three-period oscillations of AC magnetic field, into $N_{\rm vn}$ moments with equivalent intervals, which are referred to as virtual nodes [Fig.~\ref{Fig05}(c)]. We simulate time evolutions of the magnetizations $\bm m_i$ by solving the LLG equation and trace the time profile of site-averaged $z$-component magnetization $M_z(T, \tau)$ inside the readout node. Here the average is taken over sites within the circular area of detector, and the integer variable $\tau(=1, 2, \cdots, N_{\rm vn})$ is an index of the virtual nodes. As a position of the single readout node, we examine nine circular areas as its candidates to investigate possible dependence of the performance on the position of readout node. The candidate positions are labeled by 1 to 9 as shown in Fig.~\ref{Fig05}(b). Note that the position labeled by 2 is assigned to both the input node and a readout node. The radii of the circular detector areas are all 3 in units of the lattice constant. The center of Detector 5 is located at $(x,y)=(64.5, 32.5)$, and the centers of neighboring detectors are separated by distances of 58 and 26 in units of the lattice constant along the $x$ and $y$ axes, respectively.

The reservoir computing procedures for the STM and PC tasks are as follows [see also Fig.~\ref{Fig05}(d)]. We first construct the reservoir-state vector $\bm x(T)$ for each time step $T$. The vector $\bm x(T)$ is composed of signals $M_z(T, \tau)$ measured at the readout node during the time interval from $T$ to $T+1$ as,
%%%%%%%%%%%%%%%%%%%%%%%%%%%%%%%%%%%%%%%%%%%%%%%%%%%%%%%%%
\begin{align}
\bm x(T)=
\begin{pmatrix}
x_0\\
M_z(T,1)\\
M_z(T,2)\\
\cdots \\
M_z(T, N_{\rm vn})\\
\end{pmatrix},
%(x_0, M_z(T,1), M_z(T,2), \cdots, M_z(T, N_{\rm vn}))^{\rm t},
\label{STMPCrsv}
\end{align}
%%%%%%%%%%%%%%%%%%%%%%%%%%%%%%%%%%%%%%%%%%%%%%%%%%%%%%%%%
with
%%%%%%%%%%%%%%%%%%%%%%%%%%%%%%%%%%%%%%%%%%%%%%%%%%%%%%%%%
\begin{align}
M_z(T,\tau)
=\frac{1}{N_{\rm detect}}
\sum_{i\in \text{detector}} m_{z,i}(T,\tau),
\end{align}
%%%%%%%%%%%%%%%%%%%%%%%%%%%%%%%%%%%%%%%%%%%%%%%%%%%%%%%%%
where $\tau$ is the index of virtual nodes, and $N_{\rm detect}(=32)$ is the number of sites inside the area of detector. A constant bias $x_0=1$ is included as the first component of the vector $\bm x(T)$. Thus the dimension of $\bm x(T)$ is $N_{\rm vn}+1$.
%Note that for the STM and PC tasks, only a single readout node is used, and the reservoir-state vector is constructed with the measured signals at the node. The nine circles labeled by numbers 1 to 9 in Fig.~\ref{Fig05}(a) indicate its candidate positions. We investigate the figure-of-merit through varying the position of detector to examine possible dependence of the performance on the detector position.
The constructed reservoir-state vectors are translated to the output $y_{\rm out}(T,T_{\rm delay})$ for a given delay $T_{\rm delay}$ by a linear transformation with the weight vector ${\bm W}_{\rm out}(T_{\rm delay})$ as,
%%%%%%%%%%%%%%%%%%%%%%%%%%%%%%%%%%%%%%%%%%%%%%%%%%%%%%%%%
\begin{align}
&y_{\rm out}(T,T_{\rm delay})={\bm W}_{\rm out}(T_{\rm delay}) \cdot \bm x(T) 
\nonumber \\
&\hspace{1.1cm}
=W_0 + W_1 M_z(T,1) + W_2 M_z(T,2)
\nonumber \\
&\hspace{3.5cm}
+ \cdots +W_{N_{\rm vn}} M_z(T,N_{\rm vn}).
\end{align}
%%%%%%%%%%%%%%%%%%%%%%%%%%%%%%%%%%%%%%%%%%%%%%%%%%%%%%%%%
where
%%%%%%%%%%%%%%%%%%%%%%%%%%%%%%%%%%%%%%%%%%%%%%%%%%%%%%%%%
\begin{align}
&{\bm W}_{\rm out}(T_{\rm delay})
=\left(W_0(T_{\rm delay}), W_1(T_{\rm delay}), \cdots, W_{N_{\rm vn}}(T_{\rm delay})\right).
\end{align}
%%%%%%%%%%%%%%%%%%%%%%%%%%%%%%%%%%%%%%%%%%%%%%%%%%%%%%%%%
Note that the weight vectors ${\bm W}_{\rm out}(T_{\rm delay})$ differ among the delay times $T_{\rm delay}$ and among the nine detector positions. They are required to be trained independently for each delay time and for each detector position.

Figures~\ref{Fig06}(a)-(i) show time profiles of the dynamical magnetizations measured at Detectors 1-9 as responses to the first 30 random sequence of 1/0 input digits, which are respectively entered via the input node as on/off of the AC-field pulse. The unit pulse consists of three-period oscillations of AC magnetic field, duration of which corresponds to 152 time steps in the LLG integration. The horizontal axis represents the integer time $T$ in units of the pulse width. The left vertical axes represent the on-off of AC-field pulse for 1/0 input digits. The right vertical axes represent rescaled magnetization $\Delta M_z(t)$ measured at each detector, which is defined by $\Delta M_z(t) \equiv [M_z(t) - M_z^{\rm min}] \times 10^5$. Here $M_z^{\rm min}$ is the minimum value of $M_z(t)$ in initial 380 pulse units, and $M_z(t)\equiv \sum_i m_{zi}(t)/N_{\rm detect}$ is the magnetization averaged over the detector area.

In Fig.~\ref{Fig06}(b) for Detector 2, each shaded unit bar contains three oscillations of $\Delta M_z(t)$ because Detector 2 is also the input node to which the AC-field pulse is applied, and the magnetizations in this area respond instantaneously to the applied AC magnetic field. In Figs.~\ref{Fig06}(a) and (c) for Detectors 1 and 3, the magnetizations $\Delta M_z(t)$ also respond almost instantaneously to the on/off of AC-field pulses because locations of these detectors are also close to the input node (Detector 2). On the contrary, in Figs.~\ref{Fig06}(d)-(i) for distant Detectors 4-9, the magnetization oscillations are weaker in amplitude and exhibit certain delays with respect to the timing of input. For example, in the first interval from $T=0$ to $T=1$, the magnetizations start oscillating only near the end of this time interval, even though the AC-field pulse is applied to the input node (Detector 2) from the beginning in this time interval. It is also noteworthy that overall envelopes of the magnetization oscillations at these distant detectors are less correlated with the input sequence compared to those at the close detectors to the input, i.e., Detectors 1-3.

The analyses of the STM [PC] task are done as follows. The target $y_{\rm target}(T,T_{\rm delay})$ in Eq.~(\ref{STM}) [Eq.~(\ref{PC})] and the output $y_{\rm out}(T,T_{\rm delay})$ are plugged into the formula of MSE in Eq.~(\ref{MSE}) and we optimize the weight matrix for respective detectors by the pseudoinverse-matrix method using a sequence of input data for training. After this training procedure, a subsequent sequence of input data for testing are entered to investigate the performances of reservoir on the STM and PC tasks. 

In Figs.~\ref{Fig07}(a)-(d), we compare the profiles of output $y_{\rm out}(T,T_{\rm delay})$ (blue dashed lines) with those of target $y_{\rm target}(T,T_{\rm delay})$ (red solid lines) for the STM task for several choices of delay $T_{\rm delay}$, i.e., (a) $T_{\rm delay}$=1, (b) $T_{\rm delay}$=2, (c) $T_{\rm delay}$=7, and (d) $T_{\rm delay}$=10. Here the profiles of targets are calculated using Eq.~(\ref{STM}), and the measurements of readout signals are performed by setting the number of virtual nodes as $N_{\rm vn}$=62. We find that the outputs perfectly reproduce the targets when the delay is as small as $T_{\rm delay}$=1 and $T_{\rm delay}$=2 in Fig.~\ref{Fig07}(a) and Fig.~\ref{Fig07}(b), respectively. On the contrary, when $T_{\rm delay}$ is relatively large as $T_{\rm delay}$=7 and $T_{\rm delay}$=10, the discrepancies are pronounced although the tendencies of targets are reproduced to some extent [Figs.~\ref{Fig07}(c) and \ref{Fig07}(d)].

We also compare the profiles of output $y_{\rm out}(T,T_{\rm delay})$ with those of target $y_{\rm target}(T,T_{\rm delay})$ for the PC task in Figs.~\ref{Fig08}(a)-(c) for selected delays $T_{\rm delay}$, i.e., (a) $T_{\rm delay}$=1, (b) $T_{\rm delay}$=2, and (c) $T_{\rm delay}$=7. Here the profiles of target are calculated using Eq.~(\ref{PC}), and the measurements of readout signals are again performed by setting $N_{\rm vn}$=62. Figure~\ref{Fig08}(a) shows perfect coincidence between the outputs and the targets when $T_{\rm delay}$=1. However, discrepancy appears even when the delay is as small as $T_{\rm delay}$=2 [Fig.~\ref{Fig08}(b)], and they become more significant when $T_{\rm delay}$=7 [Fig.~\ref{Fig08}(c)].

To quantitatively evaluate the performances, we use the standard squared correlation $\text{Corr}^2$ between the targets and the outputs defined by~\cite{Jaeger2001,Furuta2018,Kanao2019},
%%%%%%%%%%%%%%%%%%%%%%%%%%%%%%%%%%%%%%%%%%%%%%%%%%%%%%%%%%%%%
\begin{eqnarray}
\text{Corr}^2(T_{\rm delay})=\frac{\text{Cov}^2[y_{\rm target}(T,T_{\rm delay}),y_{\rm out}(T,T_{\rm delay})]}{\text{Var}[y_{\rm target}(T,T_{\rm delay})]\text{Var}[y_{\rm out}(T,T_{\rm delay})]},\nonumber\\
\end{eqnarray}
%%%%%%%%%%%%%%%%%%%%%%%%%%%%%%%%%%%%%%%%%%%%%%%%%%%%%%%%%%%%%
with
%%%%%%%%%%%%%%%%%%%%%%%%%%%%%%%%%%%%%%%%%%%%%%%%%%%%%%%%%%%%%
\begin{eqnarray}
&&\text{Cov}[A(T),B(T)]=\frac{1}{N_T}\sum_T (A(T)-\bar{A})(B(T)-\bar{B}),
\nonumber\\
&&\text{Var}[A(T)]=\frac{1}{N_T}\sum_T (A(T)-\bar{A})^2,
\end{eqnarray}
%%%%%%%%%%%%%%%%%%%%%%%%%%%%%%%%%%%%%%%%%%%%%%%%%%%%%%%%%%%%%
where Cov and Var denote the covariance and variance, respectively, $\bar{A}$ is the average of $A(T)$ over all $T$, and $N_T$ is the number of time steps $T$. The standard squared correlation $\text{Corr}^2$ takes a value within a range of [0,1], and a larger value indicates better coincidence of the outputs with the targets. In the present analysis, this quantity is calculated using a sequence of 90 binary digits for testing after optimizing the weight matrix using a sequence of 350 binary digits for training. After calculating Corr$^2$ as a function of $T_{\rm delay}$, we take their summation over a specific range of $T_{\rm delay}$ to evaluate a quantity called capacity $C$ for both the STM and PC tasks~\cite{Jaeger2001,Furuta2018,Kanao2019},
%%%%%%%%%%%%%%%%%%%%%%%%%%%%%%%%%%%%%%%%%%%%%%%%%%%%%%%%%%%%%
\begin{eqnarray}
C = \sum^{T_{\rm delay}^{\rm max}}_{T_{\rm delay}=0}\text{Corr}^2(T_{\rm delay}).
\label{Capacity}
\end{eqnarray}
%%%%%%%%%%%%%%%%%%%%%%%%%%%%%%%%%%%%%%%%%%%%%%%%%%%%%%%%%%%%%
A larger capacity indicates that a larger amount of memory or nonlinearity is stored in the current reservoir state. Therefore, the capacity can quantify the performances of reservoir for the STM and PC tasks.

Figure~\ref{Fig09} presents the calculated squared correlations $\text{Corr}^2$ for the STM and PC tasks as functions of the delay $T_{\rm delay}$ for Detector 7. Here the examinations of both tasks are done by setting the number of virtual nodes as $N_{\rm vn}=62$. The $\text{Corr}^2$ for the STM task takes nearly unity from $T_{\rm delay}$=0 to $T_{\rm delay}$=6, in accordance with the nearly perfect coincidence between the outputs and the targets when $T_{\rm delay}=1$ [Fig.~\ref{Fig07}(a)] and $T_{\rm delay}$=2 [Fig.~\ref{Fig07}(b)]. However, it starts decreasing from $T_{\rm delay}$=7 and gradually decays as $T_{\rm delay}$ increases in accordance with the apparent discrepancy between the outputs and the targets at $T_{\rm delay}$=7 [Fig.~\ref{Fig07}(c)] and $T_{\rm delay}$=10 [Fig.~\ref{Fig07}(d)]. The $\text{Corr}^2$ for the STM task finally vanishes around $T_{\rm delay}=22$. We note that this fading memory property (also known as the echo-state property~\cite{Jaeger2001}) is essential for a workable reservoir, since it indicates that the reservoir states can be independent of the initial configuration of the physical system, after the injection of a long enough input sequence~\cite{Tanaka2019}. On the other hand, The $\text{Corr}^2$ for the PC task shows much steeper decrease. Namely, it takes nearly unity only at $T_{\rm delay}$=0 and $T_{\rm delay}$=1 but becomes suppressed abruptly ($\sim$0.8 at $T_{\rm delay}$=2 and $\sim$0.2 at $T_{\rm delay}$=3). This is consistent with the considerable discrepancy between the outputs and the targets even at $T_{\rm delay}$=2 in Fig.~\ref{Fig08}(b). The $\text{Corr}^2$ for the PC task vanishes at $T_{\rm delay}$=4.

Because the squared correlations $\text{Corr}^2$ for the STM and PC tasks decay as $T_{\rm delay}$ increases and almost vanish above $T_{\rm delay}$=22 and $T_{\rm delay}$=4, respectively [Fig.~\ref{Fig09}], we set $T_{\rm delay}^{\rm max}$=30 in Eq.~(\ref{Capacity}) to evaluate the capacities for these tasks. Figure~\ref{Fig10} shows the capacities for both STM and PC tasks, $C_{\rm STM}$ and $C_{\rm PC}$, as functions of the number of virtual nodes $N_{\rm vn}$ for respective detectors. For all the detectors, both $C_{\rm STM}$ and $C_{\rm PC}$ increase nearly monotonically as $N_{\rm vn}$ increases because larger $N_{\rm vn}$ endows more degrees of freedom in the weight matrix to describe the output data. Both capacities are saturated to certain values. The largest $C_{\rm STM}$ and $C_{\rm PC}$ are $\sim$12.8 and $\sim$2.7, respectively. These values are comparable with that of other previously proposed spintronics reservoirs under similar number of virtual nodes~\cite{Furuta2018,Kanao2019,Yamaguchi2020}, which clearly demonstrate that the magnetic skyrmion system is promising for application to physical reservoir.

%%%%%%%%%%%%%%%%%%%%%%%%%%%%%%%%%%%%%%%%%%%%%%%%%%%%%%%%%%%%%
\begin{figure}[tb]
\centering
\includegraphics[scale=1.0]{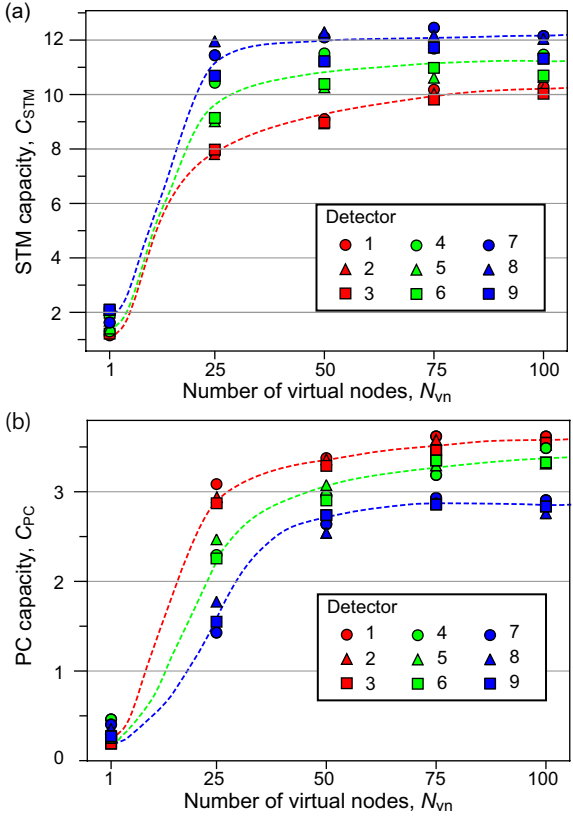}
\caption{(a) $N_{\rm vn}$-dependence of the STM capacity $C_{\rm STM}$ for respective detectors, where $N_{\rm vn}$ is the number of virtual nodes. (b) $N_{\rm vn}$-dependence of the PC capacity $C_{\rm PC}$. Detectors 1-9 are located at positions shown in Fig.~\ref{Fig05}(b). Colors of the symbols categorize these detectors in terms of the distance from the input node. Specifically, the red color is assigned to Detectors 1-3 located near the left edge in Fig.~\ref{Fig05}(b) close to the input node, whereas the blue color is assigned to Detectors 7-9 located near the right edge distant from the input node. The green color is assigned to Detectors 4-6 located in the middle of the system where the distances from the input node are intermediate. Dashed lines are guides for eyes.}
\label{Fig11}
\end{figure}
%%%%%%%%%%%%%%%%%%%%%%%%%%%%%%%%%%%%%%%%%%%%%%%%%%%%%%%%%%%%%
To investigate possible detector-position dependence of the performance, in Fig.~\ref{Fig11} we compare the capacities $C_{\rm STM}$ and $C_{\rm PC}$ for different detectors by plotting the data for specified numbers of virtual nodes, i.e., $N_{\rm vn}$=1, 25, 50, 75, and 100, extracted from Fig.~\ref{Fig10}. We use symbols of red, green, and blue colors to represent Detectors 1-3, 4-6, and 7-9, respectively, which are categorized in terms of their positions. Specifically, Detectors 1-3 represented by the red symbols are located near the left edge of the rectangular-shaped system and thus are close to or exactly on the input node, whereas Detectors 7-9 represented by the blue symbols are located near the right edge and thus are far from the input node. Detectors 4-6 represented by the green symbols are located in the middle of the system with intermediate distances from the input node. We find that there are apparent dependencies on the detector position for both $C_{\rm STM}$ and $C_{\rm PC}$, but their tendencies are opposite. 

The plots in Fig.~\ref{Fig11}(a) show that the capacities $C_{\rm STM}$ abruptly increase as $N_{\rm vn}$ increases from $N_{\rm vn}$=1 to $N_{\rm vn}$=25 and saturate to certain constant values above $N_{\rm vn}$=50 for all the detectors. We also find that there is an apparent trend that the detectors distant from the input node exhibit better performances on the STM task as indicated by larger values of $C_{\rm STM}$. Indeed, Detectors 7-9 located near the right edge (blue symbols) tend to have larger $C_{\rm STM}$, while Detectors 1-3 located near the left edge (red symbols) tend to have smaller $C_{\rm STM}$. 

On the other hand, the capacities $C_{\rm PC}$ show similar abrupt increase and saturating behavior in Fig.~\ref{Fig11}(b). However, the trend of the detector-position dependence is opposite to that of $C_{\rm STM}$. Apparently, the detectors closer to the input node tend to exhibit better performance with larger $C_{\rm PC}$ for the PC task and thus can achieve more significant nonlinearity in response. Indeed, Detectors 1-3 (red symbols) have larger $C_{\rm PC}$ than Detectors 7-9 (blue symbols).

The observed opposite trends of the detector-position dependence between the two capacities $C_{\rm STM}$ and $C_{\rm PC}$ seem to be consistent with an empirical law of the memory-nonlinearity trade-off relation in dynamical models~\cite{Dambre2012,Inubushi2017}. It was argued that the memory capacity seems to be degraded by introducing nonlinearity into the dynamics in reservoirs and vice versa. Clarification of possible connection of the magnetic skyrmion system with the dynamical systems are left for future studies.

%%%%%%%%%%%%%%%%%%%%%%%%%%%%%%%%%%%%%%%%%%%%%%%%%%%%%%%%%%%%%
\begin{figure}[tb]
\centering
\includegraphics[scale=1.0]{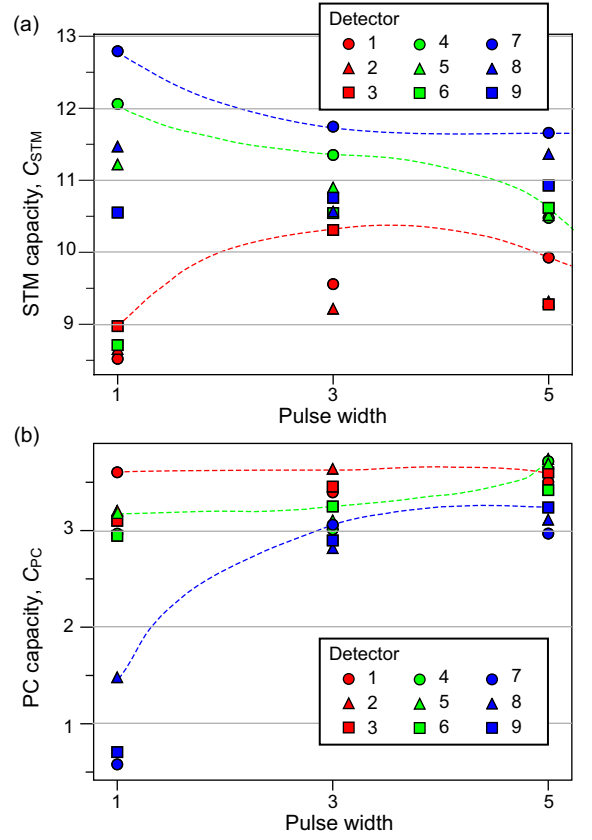}
\caption{Pulse-width dependence of (a) STM capacity $C_{\rm STM}$ and (b) PC capacity $C_{\rm PC}$. The pulse width in $x$-axis is in units of the period of AC magnetic field, which is roughly $2\pi/0.12416\approx50.6$. For pulse widths equal to 1, 3, and 5, we take integer time steps of 51, 152, and 254, respectively, for the LLG integration. Detectors 1-9 are located at positions shown in Fig.~\ref{Fig05}(b). The symbols are the same as those in Fig.~\ref{Fig11}. Dashed lines are guides for eyes.}
\label{Fig12}
\end{figure}
%%%%%%%%%%%%%%%%%%%%%%%%%%%%%%%%%%%%%%%%%%%%%%%%%%%%%%%%%%%%%
We have investigated the performances of the skyrmion spin-wave reservoir on the STM and PC tasks by taking the three-period AC-field pulse as an input unit. To study possible pulse-width dependence of the performances, we calculate the capacities $C_{\rm STM}$ and $C_{\rm PC}$ using pulses with different widths, i.e., single-period and five-period AC-field pulses. We employ the same random sequence of binary inputs as that used in the examination with the three-period AC-field pulses. For this comparison we use initial 150 binary data for training and subsequent 50 binary data for testing. The capacities $C_{\rm STM}$ and $C_{\rm PC}$ are again calculated by summing up the squared correlations Corr$^2$ in the range of $0 \leq T_{\rm delay} \leq 30$. 

Figures~\ref{Fig12}(a) and (b) show the calculated pulse-width dependencies of $C_{\rm STM}$ and $C_{\rm PC}$, respectively. Here the number of virtual nodes is fixed at $N_{\rm vn}=50$. The colored symbols are assigned to respective detectors depending on the distance from the input node in the same fashion as in Fig.~\ref{Fig11}. According to this figure, the pulse-width dependence also exhibits some characteristic behaviors. For the capacity $C_{\rm STM}$, the detectors relatively distant from the input node, i.e., Detectors 4-7 (green and blue symbols) tend to exhibit larger values than the close detectors, i.e., Detectors 1-3. In particular, Detector 7 (blue circles) exhibits the largest value of $C_{\rm STM}^{\rm max} \sim 13$, and Detector 4 (green circles) exhibits the second largest value of $C_{\rm STM}^{\rm max} \sim 12$ when the pulse width is one period of the AC magnetic field. However, the values for these two detectors decrease as the pulse width increases, indicating that we can obtain a better performance on the STM task with a shorter pulse and a detector distant from the input node. 

However, we should note that detectors located distant from the input node do not necessarily exhibit high performances on the STM task. For example, Detectors 8 and 9 (blue triangles and squares) exhibit lower or comparable values of $C_{\rm STM}$ than Detectors 4 and 5 at the intermediate distance when the pulse widths are one and three periods. Moreover, Detector 6 exhibits almost the lowest values when the pulse width is one period despite it is located at the intermediate distance. We realize that both Detectors 6 and 9 with low performances are located near the bottom edge of the rectangular system, whereas both Detectors 4 and 7 with high performances are located near the upper edge. It is known that the spin waves in magnetic skyrmions are subject to emergent magnetic fields generated by the magnetizations of topological skyrmion textures, which cause their transverse propagation called topological magnon Hall effect. Because of this effect, the propagation of spin waves in the skyrmion spin-wave reservoir can be directional and their amplitude distribution can be asymmetric, which may lead to the observed distinct performances between detectors near the upper edge and those near the bottom edge.

On the other hand, the plots of $C_{\rm PC}$ in Fig.~\ref{Fig12}(b) show no significant dependence on the pulse width for Detectors 1-3 (red symbols) and Detectors 4-6 (green symbols). The values are almost constant to be $C_{\rm PC}=$3-3.6 irrespective of the pulse width for these detectors. On the contrary, the values are considerably small as $C_{\rm PC}=$0.5-1.5 for Detectors 7-9 (blue symbols) when the pulse width is short as one period. The performances of Detectors 7 and 9 for the PC task is remarkably low ($C_{\rm PC}\sim$0.5 when the pulse width is one period). This fact seems to be in sharp contrast to the case of the STM task discussed above. Namely, the largest capacity of the STM task ($C_{\rm STM}\sim$13) is achieved by Detector 7 when the pulse width is one period. The contrasting performances for Detector 7 between the STM and PC tasks seem to be consistent with the memory-nonlinearity trade-off relation in dynamical models again.

\section{Conclusion}
In this paper, we have proposed a concept of the skyrmion spin-wave reservoir and have examined its properties and performances. We have investigated three of the required characteristics of reservoir, i.e., the generalization ability, the short-term memory function, and the nonlinearity of our skyrmion spin-wave reservoir by imposing three standard tasks, i.e., the duration-estimate task, the short-term momory task, and the parity-check task. Through these investigations, we have demonstrated that the skyrmion spin-wave reservoir possesses high abilities for information processing. Importantly, magnetic skyrmions emerge spontaneously in magnetic specimens with broken spatial inversion symmetry under application of static magnetic field via self-organization process. Therefore, the proposed skyrmion reservoir requires neither advanced nanofabrication nor complicated manufacturing for their production in contrast to other previously proposed magnetic reservoirs with elaborate spintronics devices, e.g., spin-torque oscillators and magnetic tunnel junctions. Our proposal will necessarily pave a way to the realization of practically useful spintronics reservoirs of high performance.

\section{Acknowledgement}
This work is supported by Japan Society for the Promotion of Science KAKENHI (Grant No. 20H00337), CREST, the Japan Science and Technology Agency (Grant No. JPMJCR20T1), and a Research Grant in the Natural Sciences from the Mitsubishi Foundation.

\end{document}